\documentstyle[11pt,epsfig]{article}

\def\laq{\raise 0.4ex\hbox{$<$}\kern -0.8em\lower 0.62
ex\hbox{$\sim$}}
\def\gaq{\raise 0.4ex\hbox{$>$}\kern -0.7em\lower 0.62
ex\hbox{$\sim$}}

\begin{document}

\begin{titlepage}

\begin{flushright}
CERN-PH-TH/2004-173
\end{flushright}

\vspace*{1.8 cm}

\begin{center}

{\huge{Magnetized initial conditions for CMB anisotropies}}

\vspace{1cm}

\large{Massimo Giovannini}

\bigskip
\normalsize

\vspace{0.3cm}
{\sl Department of Physics, Theory Division, CERN, 1211 Geneva 23, Switzerland }

\vspace*{5mm}

\begin{abstract}
This paper introduces a systematic treatment of the linear theory 
of scalar gravitational perturbations in the presence of a fully inhomogeneous 
magnetic field. The analysis is conducted both in the synchronous and in the 
conformally Newtonian gauges. The cosmological plasma is assumed to be composed  of 
cold dark mattter, baryons, photons,  neutrinos.
The problem of super-horizon initial conditions for the 
fluid variables of the various species and for the coupled system 
of Boltzmann-Einstein equations is discussed  
in the presence of an inhomogeneous magnetic field.  
The tight coupling approximation for the Boltzmann hierarchy 
is extended to the case where  gravitating magnetic fields are included.
\end{abstract}

\end{center}

\end{titlepage}

\renewcommand{\theequation}{1.\arabic{equation}}
\section{Introduction}
\setcounter{equation}{0}

The impact of magnetic fields on the anisotropies of the Cosmic Microwave 
Background (CMB) has been a subject of active investigation since the early 
attempts of Zeldovich 
\cite{zel1,zel2}, where it was proposed  that all the anisotropy in the CMB 
could be generated by a magnetic field. 
Following the development of inflationary cosmology it is now clear that  the 
explanation of the large-scale temperature anisotropies 
should be attributed to some adiabatic (or quasi-adiabatic) 
mode, which was present outside the horizon prior to the decoupling of 
radiation from matter.

The theory of scalar gravitational fluctuations is a well developed subject 
and it is an essential tool setting  the initial conditions for the 
evolution of CMB anisotropies. 
When initial conditions are set, usually deep within the radiation 
era (but after neutrino decoupling taking place around $1$ MeV),  
the dominant component of the plasma are photons and 
(effectively massless) neutrinos. 
The subdominant 
component of the plasma is formed by  baryons, electrons, 
cold dark matter (CDM) particles.
CDM particles are only coupled to gravitational interactions and behave 
like a perfect relativistic fluid. Also the photons, leptons and  baryons, 
being tightly coupled through Thompson scattering, behave like perfect 
relativistic fluids.
On the contrary, neutrinos  are essentially collisionless and, therefore, 
do not really behave like a perfect fluid. Because of this physical 
difference, neutrinos should be described through 
the appropriate Boltzmann hierarchy of their phase-space distribution. 
If the dark energy is paramertized in terms of a cosmological term, the 
gravitational fluctuations of this sector do not affect the problem 
of the initial conditions.

If primordial fluctuations of the geometry are present outside 
the horizon before matter-radiation equality, they imprint 
also fluctuations in the density contrasts and peculiar velocities of the different 
species. 
The quantitative evolution of the various plasma  quantities is  determined by the fluctuations 
of the geometry by solving the coupled system of Einstein and fluid equations 
in the radiation-dominated epoch when the relevant modes are still outside the horizon. 
Following  the classification scheme 
pioneered by Bardeen \cite{bardeen}, 
the initial conditions may be classified into adiabatic or isocurvature modes. 
In the adiabatic case the total fluctuations in the 
entropy density of the CDM--photon--lepton--baryon fluid vanish at large distance scales
which 
is compatible with a constant mode of curvature fluctuations outside the horizon. In the 
case of isocurvature modes, in the usual terminology, the fluctuations in the entropy density do not vanish;
according to the linear analysis of the Einstein-fluid system,  this is compatible, in some cases,
 with non-constant modes of the curvature fluctuations. In spite of this 
simplified  classification, the situation may be more complicated since different isocurvature modes may be
allowed for the different species present in the plasma. Moreover, it is also 
quite plausible to have a situation where the initial conditions are, predominantly, of adiabatic nature, 
but with a subdominant isocurvature component.

Although an excellent approximation early on, the tight-coupling assumption breaks down at later times, when 
photons and baryons decouple. In spite of this caveat, a systematic use of  the tight-coupling expansion 
allows important analytical estimates of the produced CMB anisotropies \cite{HS1}.

In the theory of the CMB anisotropies, 
the problem of initial conditions is often presented in two complementary descriptions, namely the conformally 
Newtonian and the synchronous gauges. Both gauges are quite  useful for different reasons (see also \cite{MB}). 
The conformally Newtonian gauge (often dubbed  longitudinal) is effective  in the discussion of the
 evolution of the fluctuations while they are still outside the horizon. However, in such a gauge, the
discussion of the initial conditions for the CMB anisotropies may be rather complicated. For instance, 
there are physical isocurvature modes that are singular in the longitudinal gauge but not in the 
synchronous gauge. Furthermore, various codes needed for the numerical calculations 
of the CMB spectra are formulated within the synchronous coordinate system. There have been, in the 
past, controversies concerning the use of the synchronous gauge \cite{PV1}. 
The main caveat was that gauge modes may appear in the 
synchronous gauge. These gauge modes, however, can be precisely disentangled from the physical 
ones \cite{PV1}, so that
a pragmatic  approach envisaged some time ago \cite{MB} has been to discuss the problem of initial conditions 
in both gauges in parallel.

Suppose  now that inhomogeneous magnetic fields are present. In this case two major effects can be envisaged:
\begin{itemize}
\item{} inhomogeneous magnetic fields can gravitate, i.e. they can affect the perturbed Einstein equations, thus
becoming sources of the fluctuations of the geometry;
\item{} inhomogeneous magnetic fields can impart different velocity gradients to the 
baryon--photon--lepton fluid, but not to the neutrinos and to the CDM components.
\end{itemize}
Deep within the radiation epoch, large-scale magnetic fields can be treated, in 
first approximation, as interacting with a single globally neutral fluid. 
In fact the two electrically charged species present in the problem are the electrons and the baryons which 
are in thermal equlibrium at a common temperature. The typical length scales of interest for the present 
discussion are the ones 
much larger than the typical magnetic diffusivity scale (set by the finite value of the 
conductivity) and also much larger than the Silk damping scale (set by the finite value of the thermal diffusivity scale).
Furthermore, for this range of scales we are not interested in the propagation 
of high frequency electromagnetic waves whose specific analysis would clearly require 
a two-fluid plasma treatment. The one-fluid plasma description if often dubbed as magnetohydrodynamics (MHD).
In MHD, the current density, the magnetic field and the electric field are all solenoidal 
and the Lorentz force affecting the baryon peculiar velocity modifies the standard tight-coupling approximation. 

By looking at the evolution of the brightness of the temperature fluctuations it is 
rather clear that, to zeroth order in the tight-coupling expansion, the monopole and dipole equations are 
modified. It is known \cite{HS1} that to  zeroth order in the tight coupling expansion the CMB is not linearly 
polarized since, to this order, photons and baryons are so strongly coupled that the photon distribution is isotropic in the 
baryon rest frame. The photon distribution being isotropic, Thompson scattering does not polarize the CMB.
To first order in the tight-coupling expansion the polarization of the CMB is proportional 
to the quadrupole of the photon distribution. The quadrupole in the temperature fluctuations is, in turn, produced by the free 
streaming of the dipole between collisions. If a Lorentz force term is present, the evolution of the monopole acquires a further 
source term and, as a consequence, all the higher orders in the tight coupling expansion are modified.

Various analyses of the possible impact of magnetic fields on standard CMB physics\footnote{By standard 
CMB physics we mean that a constant mode of adiabatic (or quasi-adiabatic) fluctuations is assumed to be present outside the 
horizon prior to equality.} have been proposed up to now (see, for instance, \cite{TK1} for a short review on this specific subject). 
In \cite{koh} the analysis of scalar fluctuations of the geometry is made under the assumption that 
the Lorentz force vanishes. In MHD this assumption translates into a specific class of magnetic field configurations. 
However, even in the case of a force-free configuration the specific 
initial conditions used by the authors in order to compute the CMB anisotropies are not specifically discussed, i.e. 
no solution for super-horizon sized fluctuations including large-scale magnetic fields, is presented. Furthermore, no 
specific investigation of the impact of gravitating magnetic fields on the tight-coupling expansion has been attempted.
In \cite{mack} an interesting analysis of the impact of large-scale magnetic fields on CMB anisotropies 
has been performed, mainly for the vector and tensor modes of the geometry. The scalar fluctuations have not been discussed.
In \cite{lewis} attention is always paid to vector and tensor modes and the problem of initial conditions 
is more accurately specified. It is clear that without a precise specification of the initial 
conditions for the CMB anisotropies the numerical analysis is rather difficult to implement
since it is unclear what should be, for instance, the initial values of the lowest multipoles 
of the phase-space distribution for the various species. The need of an accurate analysis 
of the problem of initial conditions can be also understood by observing that 
the system of perturbed equations, even if linear, has many unknowns so different 
solutions, describing different physical situation may be possible. 

The effect of large-scale magnetic fields on CMB anisotropies may also be
discussed in the case when the magnetic field is not 
{\em completely} inhomogeneous. There could be situations in which there is an homogeneous component 
of the magnetic field (for instance along the $\hat{z}$ axis. This case is similar to the one 
originally investigated in \cite{zel1,zel2}. In this framework bounds on such a magnetic field can be derived \cite{B1}.
It could also happen that the uniform magnetic field supports inhomogeneities in the bulk velocity field
affecting, consequently, the CMB anisotropies. One example in this direction are the Alfv\'en waves, whose 
effects and implications have been analysed under different  approximations by various authors \cite{adams,SB1,SB2,chen}.
The simplification of having a uniform magnetic field also allowed the analysis of possible Faraday rotation 
effect of the CMB polarization plane \cite{K1,H1,mg1,SC}.

The motivation for including a uniform magnetic field relies on the simplicity of the configuration. 
However, the magnetic fields produced in the early Universe, as far as we understand the problem, 
are unlikely to be produced in a perfectly uniform configuration. They are most likely to be 
produced either from the amplification of vacuum fluctuations of some primordial gauge field or from 
some phase transitions. In all these cases magnetic fields are expected to be fully inhomogeneous
(see, for instance, \cite{giovannini} for a recent review on the r\^ole of large-scale 
magnetic fields in cosmology and astrophysics).

The purpose of the present study is to give a systematic analysis of the theory of scalar 
gravitational fluctuations in the presence of a fully inhomogeneous magnetic field and in the 
context of the conventional Bardeen formalism.  
This analysis 
is mandatory both to derive accurate bounds on large-scale magnetic fields from CMB physics 
and  to bring the study of magnetized initial conditions to the same standard 
as in the non-magnetized case. To the best of our knowledge this problem did not receive 
specific attention in the literature concerning the linearized theory of scalar 
fluctuations in the presence of magnetic fields.  The results reported in the present paper 
are general in the sense that no specific configuration of the magnetic field 
is assumed. The only assumption, as previously stressed, is that the magnetic field 
is fully inhomogeneous. 

It is important to mention that various studies discussing the evolution 
of large-scale magnetic field exist in the covariant approach which is 
somehow complementary to the one adopted in the present paper. For 
a full discussion of the problem see \cite{ts1,ts2} and references therein.

The plan of the present paper is  the following. In Section 2 the evolution equations for gravitating 
inhomogeneous magnetic fields will be derived in the MHD approximation and in the context 
of the conformally Newtonian gauge. 
In Section 3 the initial conditions for the CMB anisotropies will be analysed. Solutions for 
super-horizon-sized fluctuations will be presented in different cases.
Section 4 is devoted to the analysis of the synchronous gauge description.
In Section 5 the tight-coupling expansion 
will be revisited in the presence of inhomogeneous 
magnetic fields.  Section 6 contains some concluding remarks. Various useful 
technical results have been collected in the appendix even though, in some cases, they were 
already present in the literature, but within a different set of conventions or within 
a different context.

\renewcommand{\theequation}{2.\arabic{equation}}
\section{Gravitating inhomogeneous magnetic fields}
\setcounter{equation}{0}

The treatment of magnetic fields in a globally neutral plasma differs considerably 
according to the specific range of scales relevant for the problem under discussion. For instance, the  
electromagnetic branch of the  spectrum of plasma excitations, in a cold or warm magnetized plasma, 
can only be partially addressed within a  two-fluid description \cite{kra,boyd}.  Partially means, in the 
present context, that the dispersion relations can be obtained but the typical damping scales 
require the solution of the full kinetic system of equations for 
the different species (the so-called Vlasov--Landau approach).

The phenomena occurring over much smaller frequencies and over large length scales can be described 
on the basis of our knowledge of terrestrial plasma, by an appropriate 
one-fluid description, which is provided by  MHD \cite{biskamp}. The main idea behind 
the MHD approximation is, in short, the following. Starting from a two-fluid 
description (for instance electron and baryon fluids) it is possible 
to define appropriate one-fluid variables. For instance the bulk velocity of the plasma 
appearing in the MHD description is, in flat space-time, 
\begin{equation}
\vec{v} = \frac{ m_{\rm b} \vec{v}_{\rm b} + m_{\rm e} \vec{v}_{\rm e}}{m_{\rm b} + m_{\rm e}}.
\label{bulkvel1}
\end{equation}
which is, in practice, the centre-of-mass velocity 
of the baryon--electron system. In the specific case touched by the present discussion, both 
baryons and electrons are non-relativistic and in thermal 
equilibrium $T_{\rm eb}$. 

The reduced MHD description \cite{biskamp} has been employed in the analysis of different problems
arising in connection with large-scale magnetic fields. For instance, the evolution of 
these fields in curved space-times is normally discussed 
within a MHD approach \cite{enq1,enq2,ol1} (see also for a review \cite{ol2,enq4}). 
More formal discussions on MHD and on two-fluid descriptions in curved space-times 
can be found in \cite{hol,GFK,GFK2}. Finally, the reduced MHD description 
has been used in the analysis of the implications of large-scale magnetic fields 
on structure formation \cite{was,col}. 

\subsection{The perturbed system of Einstein equations}
Consider now the  Friedmann--Robertson--Walker (FRW) line element
written in the conformally flat case
\begin{equation}
ds^2 = a^2(\eta)[d\eta^2 - d \vec{x}^2],
\label{metric}
\end{equation}
where $\eta$ is the conformal time coordinate.
Since the magnetic fields are fully inhomogeneous, they will not affect the homogeneous background 
whose equations are, in the spatially flat case:
\begin{eqnarray}
&&{\cal H}^2 = \frac{8\pi G}{3} a^2 \rho \sum_{\lambda} \Omega_{\lambda} ,
\label{b1}\\
&& {\cal H}^2 - {\cal H}' = 4 \pi G a^2 \rho \sum_{\lambda} \Omega_{\lambda} (1 + w_{\lambda}),
\label{b2}\\
&& \rho_{\lambda}' + 3 {\cal H} ( \rho_{\lambda} + p_{\lambda}) =0,
\label{b3}
\end{eqnarray}
where ${\cal H} = (\ln{a})'$ and the prime denotes a derivation with respect to $\eta$.
In Eqs. (\ref{b1})--(\ref{b3})  the summation index $\lambda$ refers to each component of the plasma, i.e. baryons, photons, neutrinos and CDM particles;  
$w_{\lambda} = p_{\lambda}/\rho_{\lambda}$ and $\Omega_{\lambda}$ are, respectively,  the 
 barotropic index for the given species $\lambda$ and  the energy density (in critical units) for each component 
of the fluid. In view of the  subsequent discussions, it is appropriate to define here also the usual Hubble parameter, i.e.
$H = \dot{a}/a$, where the overdot denotes a derivation with respect to the cosmic time coordinate $t$,
which is related to $\eta$ by the differential relation $ d\eta a(\eta) = dt$. In terms of these definitions 
we also have ${\cal H} = a H$.

Let us now consider the fluctuations of the  homogeneous FRW metric (\ref{metric}).
In this section the conformally Newtonian coordinate system will be discussed and, therefore, 
the  metric (\ref{metric}) can be consistently perturbed in terms of the two longitudinal 
scalar degrees of freedom $\phi$ and $\psi$ 
\begin{equation}
\delta g_{00} = 2 a^2 \phi,\,\,\,\,\,\,\,\,\,\,\,\,\,\, \delta g_{i j} = 2 a^2 \psi \delta_{i j}.
\label{longgauge}
\end{equation}

Defining the fluctuations in the Ricci tensor and in the Ricci scalar as $ \delta R_{\mu}^{\nu}$ and $\delta R$, 
the perturbed Einstein equations, 
\begin{equation}
\delta R_{\mu}^{\nu}  - \frac{1}{2}\delta_{\mu}^{\nu} \delta R = 
8 \pi G a^2 (\delta T_{\mu}^{\nu} + \delta \tau_{\mu}^{\nu} ),
\label{pert1}
\end{equation} 
relate the fluctuations of the geometry to the fluctuations of the matter sources, $\delta T_{\mu}^{\nu}$,  and
to the fluctuations of the electromagnetic energy-momentum tensor, $\delta \tau_{\mu}^{\nu}$.

In explicit terms the $(00)$, $(0i)$ and $(ij)$ components of  Eq. (\ref{pert1}) lead, respectively, to
\begin{eqnarray}
&& \nabla^2 \psi  - 3 {\cal H} ( {\cal H} \phi + \psi') = 4\pi G a^2 [ \delta T_{0}^{0}\,\, +\,\, \delta \tau_{0}^{0}],
\label{p00l}\\
&& - \partial^{i}( {\cal H} \phi + \psi') = 4 \pi G a^2 ( \delta T_{0}^{i} \,\,+\,\, \delta \tau_{0}^{i}),
\label{p0il}\\
&& \biggl[ \psi'' + {\cal H} ( 2\psi' +\phi') + ( 2 {\cal H}' + {\cal H}^2) \phi + 
\frac{1}{2} \nabla^2(\phi - \psi) \biggr] \delta_{i}^{j} 
\nonumber\\
&& - \frac{1}{2} \partial_{i}\partial^{j} ( \phi - \psi) = - 4\pi G a^2 [\delta T_{i}^{j}\,\, +\,\, \delta\tau_{i}^{j} ],
\label{pijl}
\end{eqnarray}
where 
\begin{eqnarray}
&& \delta \tau_{0}^{0} = \frac{1}{8\pi a^4} ( \vec{E}^2 + \vec{B}^2),
\label{tau00}\\
&& \delta \tau_{0}^{i} = \frac{1}{4\pi a^4} \vec{E} \times \vec{B} ,
\label{tau0i}\\
&& \delta \tau_{i}^{j} = \frac{1}{4 \pi a^4} \biggl[ E_{i} E^{j} + B_{i} B^{j} - \frac{1}{2} (\vec{B}^2 + \vec{E}^2) \delta_{i}^{j}\biggr],
\label{tauij}\\
&& \delta T_{0}^{0} = \sum_{\lambda} \rho_{\lambda} \delta_{\lambda}, 
\label{T00}\\
&& \delta T_{i}^{j} = -\sum_{\lambda} w_{\lambda} \rho_{\lambda} \delta_{\lambda} \delta_{i}^{j} + \Sigma_{i}^{j}, 
\label{Tij}\\
&& \delta T_{0}^{i} = \sum_{\lambda} ( 1 + w_{\lambda}) \rho_{\lambda} v_{\lambda}^{i},
\label{T0i}
\end{eqnarray}
where $ \delta_{\lambda} = \delta \rho_{\lambda}/\rho_{\lambda}$ and  $v^{i}_{\lambda}$ are, respectively, the 
density contrast and the peculiar velocity for each particle species; $\Sigma_{i}^{j}$  is the traceless component 
of the energy-momentum tensor of the fluid sources, i.e. 
\begin{equation}
\Sigma_{i}^{j} = \delta T_{i}^{j} - \frac{1}{3} \delta_{i}^{j} \delta T.
\end{equation}
According to the usual notation the anisotropic stress can also be written, in Fourier space, 
as 
\begin{equation}
{\cal Q} =\partial_{j}\partial^{i} \Sigma_{i}^{j} = - k^2 \sum_{\lambda}  \sigma_{\lambda} ( 1 + w_{\lambda}) \rho_{\lambda},
\label{shdef}
\end{equation}
where $\sigma_{\lambda}$ denotes the fraction contributed by each species $\lambda$ to the total 
anisotropic stress. Notice, as it will be more extensively discussed later, that the dominant source of anisotropic stress 
comes, for temperatures below the MeV, from the neutrinos.

Finally in Eqs. (\ref{tau00})--(\ref{tauij}) the components of the energy-momentum tensor have been written directly using
the rescaled electric and magnetic fields $ \vec{E}$ and $\vec{B}$, which can  also be  expressed in terms of the corresponding 
flat-space fields as
\begin{equation}
\vec{E}(\eta, \vec{x})  = a^2(\eta) \vec{{\cal E}}(\eta, \vec{x}) ,\,\,\,\,\,\,\,\,\,\,
\vec{B}(\eta, \vec{x})  = a^2(\eta) \vec{{\cal B}}(\eta, \vec{x}).
\label{rescaled}
\end{equation}
In the following the evolution of the different quantities appearing in the right-hand side of Einstein equations 
will be discussed in detail. 

\subsection{Electromagnetic fields and baryons}

The  Maxwell equations can be studied using the rescaled fields proposed in Eq. (\ref{rescaled}) 
\begin{eqnarray}
&&\vec{B}' = - \vec{\nabla}\times \vec{E},
\label{mx1}\\
&& \vec{\nabla} \times  \vec{B} = 4\pi \vec{J} + \vec{E}' ,
\label{mx2}\\
&& \vec{\nabla}\cdot \vec{E} = 4 \pi \rho_{\rm q} ,
\label{mx3}\\
&& \vec{\nabla} \cdot \vec{B} =0,
\label{mx4}
\end{eqnarray}
where $\rho_{\rm q}$ is the charge density ($\rho_{\rm q} \simeq 0$ in a globally neutral plasma);
$ \vec{J} = a^3 \vec{j}$ is the Ohmic current density 
\begin{equation}
\vec{J} = \sigma ( \vec{E} + \vec{v} \times \vec{B}). 
\label{J}
\end{equation}
In Eq. (\ref{J}), $\sigma$ is the conductivity that is related to the flat-space conductivity as 
$ \sigma = a \sigma_{c}$. Notice that in the reduced MHD description the Ohm law follows, in the reduced MHD 
approach,
by taking the difference of the equations describing the momentum 
conservation for electrons and for ions. This procedure is rather tricky and, in principle 
other terms may appear in the generalized Ohm law such as the thermoelectric term and the Hall term
\cite{kra,giovannini}. These will be neglected here. Finally, as 
pointed out after Eq. (\ref{bulkvel1}), the bulk velocity field appearing in Eq. (\ref{J}) can be identified with 
the velocity of the baryons.

The evolution for the baryon velocity field can be derived by perturbing 
 the covariant conservation equation (including  the appropriate electromagnetic contribution), i.e. 
\begin{equation}
\partial_{\mu} \delta T^{\mu\nu}_{\rm b}  + \delta \Gamma_{\mu\alpha}^{\nu} \overline{T}^{\mu\alpha}_{\rm b} + 
+\overline{\Gamma}_{\mu\alpha}^{\nu} \delta T^{\mu\alpha}_{\rm b} +\delta \Gamma_{\beta\mu}^{\mu} \overline{T}^{\beta\nu}_{\rm b}+ 
\overline{\Gamma}_{\beta\mu}^{\mu} \delta T^{\beta\nu}_{\rm b}  - F^{\nu\alpha} j_{\alpha} =0,
\end{equation}
where $\overline{T}^{\mu\nu}_{\rm b}$ and $\delta T^{\mu\nu}_{\rm b} $ are, respectively, the energy-momentum of the baryons and its 
first-order fluctuation; $F^{\nu\alpha}$ is the Maxwell field strength and $j_{\alpha}$ is 
the current density. An analogous notation is used for the Christoffel symbols computed on the background (denoted by an overline) and for 
their first-order fluctuations. Thus, defining  $\theta_{\rm b} = \vec{\nabla} \cdot \vec{v}_{\rm b}$, the evolution 
equation for the peculiar velocity of the baryons and for the density contrast reads:
\begin{eqnarray}
&& \theta_{\rm b}' = - {\cal H} \theta_{\rm b} -  c_{s}^2 \nabla^2 \delta_{\rm b} -\nabla^2 \phi + \frac{4}{3} \frac{\Omega_{\gamma}}{\Omega_{\rm b}} a n_{\rm e} 
x_{\rm e} \sigma_{\rm T} ( \theta_{\gamma} - \theta_{\rm b}) + \frac{\vec{\nabla}\cdot[\vec{J}\times\vec{B}]}{a^{4}\rho_{b}},
\label{baryon1}\\
&& \delta_{\rm b}' = 3 \psi' - \theta_{\rm b}.
\label{baryon2}
\end{eqnarray} 
where $x_{\rm e}$ is the ionization fraction of the plasma, $\sigma_{\rm T}$ the Thompson cross section; $\Omega_{\gamma}$ $\Omega_{\rm b}$ are, respectively,
the energy densities, in critical units, for photons and baryons. At this point we can also define the differential 
optical depth for Thompson scattering:
\begin{equation}
\tau' = x_{\rm e} n_{\rm e} \sigma_{\rm T} a(\eta).
\label{diffopdep}
\end{equation}

Equations (\ref{mx1})--(\ref{mx4}), together with Eqs. (\ref{J}) and (\ref{baryon1}), can be studied in the MHD approximation. 
Since the plasma is globally neutral ($\rho_{\rm q} \simeq 0$), 
the charge density of the electrons will be exactly compensated by the baryons and, according to Eq. (\ref{mx3}),  the 
electric field will be solenoidal. Moreover, for sufficiently small frequencies, the displacement current can be neglected in Eq. (\ref{mx2}). This implies that 
also the Ohmic current will be  solenoidal, i.e. 
\begin{equation}
\vec{\nabla} \cdot \vec{J} =0, \,\,\,\,\,\,\vec{J} = \frac{1}{4 \pi} \vec{\nabla} \times \vec{B}.
\end{equation}
As a result, the Ohmic electric field vanishes exactly in the limit of vanishing conductivity, i.e. since 
\begin{equation}
\vec{E} \sim \frac{\vec{J}}{\sigma} =\frac{\vec{\nabla}\times \vec{B}}{\sigma}, 
 \end{equation}
the electric fields appearing in Eqs. (\ref{tau00}) and (\ref{tauij}) lead to a contribution that is suppressed as $1/\sigma^2$. Similarly, 
the electric field appearing in Eq. (\ref{tau0i}) leads to a contribution that is suppressed as $1/\sigma$. 

Recalling, as previously defined, that $\sigma = \sigma_{c} a$, for temperatures $ T < 1~ {\rm MeV}$ the conductivity 
can be expressed as \cite{kra,giovannini} 
\begin{equation}
\sigma_{c} \simeq \frac{1}{\alpha_{\rm em}} \biggl( \frac{T_{\rm e b}}{m_{\rm e}} \biggr)^{1/2},
\end{equation}
where $T_{\rm eb}$, as previously defined,  is the common temperature of electrons and baryons.
The finite value of the conductivity sets a typical diffusion scale for the evolution 
of the magnetic fields which is typically of the order of 
\begin{equation}
L_{\sigma}(T) = \frac{\sigma_0^{-1/2} g_{\ast}^{1/4}}{72.2} \biggl( \frac{T_{\rm e b}}{M_{\rm P}}\biggr)^{1/2} 
\biggl(\frac{T_{\rm e b}}{m_{\rm e}}\biggr)^{-1/4}~~
L_{H}(T),~
\end{equation}
where $ L_{\rm \sigma}$ is the physical diffusion scale, $L_{H} \simeq H^{-1}$ is the Hubble radius  and $g_{\ast}$ the number of relativistic 
degrees of freedom. Since, in our problem ${\rm eV} < T_{\rm e b} < {\rm MeV}$, $L_{\sigma} \ll L_{H}$ which means that the finite value of
the conductivity only affects scales which are much smaller than the Hubble radius. 

Similar, but quantitatively different, remarks can be made for the thermal diffusivity \cite{giovannini}. For temperature smaller than the temperature 
of neutrino decoupling, photons are the most efficient source of momentum transfer and the thermal 
diffusivity scale can be written as 
\begin{equation}
\frac{L_{\rm diff}^{(\gamma)}(T)}{L_{H}(T)}\simeq 1.03 \times 10^{-5} 
\biggl(\frac{\Omega_{\rm b} h_0^2}{0.02}\biggr)^{-1/2} \biggl(\frac{x_{\rm e}}{0.5}\biggr)^{-1/2} \biggl(\frac{g_{\ast}}{10.75}\biggr)^{1/4}
\sqrt{\frac{{\rm MeV}}{T_{\rm e b}}},
\label{difm}
\end{equation}
where, according to previous definitions (see Eq. (\ref{baryon1}),  $x_{\rm e}$ is the ionization fraction, $\Omega_{\rm b}$ the energy density
 of baryons in critical units. 
 
\subsection{Cold dark matter}

The evolution of the CDM particles follows from the fluctuations of the covariant conservation equation whose first-order fluctuation 
leads, in Fourier space, to
\begin{eqnarray}
&&\theta_{\rm c}' + {\cal H} \theta_{c} = k^2 \phi,
\label{CDM1}\\
&& \delta_{\rm c}' = 3 \psi' - \theta_{\rm c}.
\label{CDM2} 
\end{eqnarray}
where, following the previous notation $ \theta_{\rm c} = \partial_{i} v^{i}_{\rm c} $ and 
$\delta_{\rm c} = \delta\rho_{\rm c}/\rho_{\rm c}$.

\subsection{Photons and massless neutrinos}
For the neutrinos the evolution equations follow only partially from the energy-momentum 
conservation. In fact neutrinos at early times after neutrino decoupling (occurring around $1$ MeV) 
obey the collisionless Boltzmann equation (see the appendix for a more detailed discussion).
To describe neutrinos during and after horizon crossing requires a Boltzmann hierarchy for $\delta_{\nu}$, $\theta_{\nu}$ and for 
the higher multipole moments, i.e. $\ell \geq 2$,  of the neutrino phase-space density ${\cal F}_{\nu\ell}$.
With this caveat in mind, after neutrino decoupling, at temperatures of about $1$ MeV, massless neutrinos obey, in Fourier space, 
the following set of equations (see the appendix for a swift derivation within the metric conventions adopted in Eqs. (\ref{metric}) and (\ref{longgauge}))
\begin{eqnarray}
&& \delta_{\nu}' = - \frac{4}{3} \theta_{\nu} + 4\psi',
\label{nu1}\\
&& \theta_{\nu}' = \frac{k^2}{4} \delta_{\nu} - k^2 \sigma_{\nu} + k^2 \phi,
\label{nu2}\\
&& \sigma_{\nu}' = \frac{4}{15} \theta_{\nu} - \frac{3}{10} k {\cal F}_{\nu 3},
\label{nu3}
\end{eqnarray}
where $\sigma_{\nu} = {\cal F}_{\nu 2}/2$ is the quadrupole moment of the (perturbed) neutrino phase-space distribution and, as 
introduced above, ${\cal F}_{\nu\ell}$ is the $\ell$-th multipole.

The photon and neutrino evolution equations differ in the presence of an anisotropic stress term $\sigma_{\nu}$. 
Because of their frequent scattering by charged leptons and baryons, photons are unable, at early times, to develop 
a quadrupole moment in their velocity distribution. As a consequence, when a mode enters the horizon, the photons behave as a 
perfect fluid, while the neutrinos free stream, creating inhomogeneities in the energy density, pressure and momentum density.
This is the physical reason why the anisotropic stress term appearing in Eq. (\ref{Tij}) is effectively dominated, at early 
times, by the neutrino contribution.
Also for the photons it is more appropriate to discuss the full Boltzmann hierarchy, especially 
in the light of the subsequent analysis of the tight-coupling approximation. This analysis 
will be summarized in the appendix. In the fluid approximation the 
evolution equations of the photons are given by the covariant conservation of the 
energy-momentum tensor, i.e. 
\begin{eqnarray}
&& \delta_{\gamma}' = - \frac{4}{3} \theta_{\gamma} + 4 \psi',
\label{phot1}\\
&& \theta_{\gamma}' = \frac{k^2}{4} \delta_{\gamma} + k^2 \phi  + a n_{\rm e} \sigma_{\rm T} ( \theta_{\rm b} - \theta_{\gamma} ).
\label{phot2}
\end{eqnarray}
At early times the characteristic time for the synchronization of the photon and baryon velocities is 
$\tau_{{\rm b}\gamma} \sim (n_{\rm e} \sigma_{\rm T})^{-1}$, which is small with respect to the expansion time and to the oscillation period. 
Thus the tight-coupling between photons and baryons means that, in the first approximation,  $\sigma_{\rm T} \to \infty$ and
$ \theta_{\gamma}\simeq \theta_{\rm b}$. A more detailed  analysis  of the tight coupling expansion is postponed to Section 5.

\renewcommand{\theequation}{3.\arabic{equation}}
\section{Large-scale solutions}
\setcounter{equation}{0}

The full system of  scalar fluctuations 
in the longitudinal gauge will now be consistently discussed and solved. 
From Eq. (\ref{p00l}) we obtain, in Fourier space\footnote{In order to avoid possible confusions 
with subscripts referring to the different species of the plasma we will avoid to introducing a further subscript labelling the Fourier mode.}
 and using  Eq. (\ref{b1}):
\begin{equation}
- 3 {\cal H} ( {\cal H} \phi + \psi') - k^2 \psi = \frac{3}{2} {\cal H}^2 [ ( R_{\nu} \delta_{\nu} + ( 1 - R_{\nu}) \delta_{\gamma}) + \Omega_{\rm B}(k) + 
\Omega_{\rm b} \delta_{\rm b} + \Omega_{\rm c} \delta_{\rm c}],
\label{p00lex}
\end{equation}
where, for $N_{\nu}$ species of massless neutrinos,   
\begin{equation}
R= \frac{7}{8} N_{\nu} \biggl( \frac{4}{11}\biggr)^{4/3},\,\,\,\,\,\,\, R_{\nu}  = \frac{R}{1 + R},\,\,\,\,\,\,\,\, R_{\gamma} = 1 - R_{\nu},
\end{equation}
so that $R_{\nu}$ and $R_{\gamma}$ represent the fractional contributions of photons and neutrinos to the total density at early times 
deep within the radiation-dominated epoch.

In Eq. (\ref{p00lex}) the contribution of the magnetic energy density has been parametrized, in Fourier space, as  
\begin{eqnarray}
\Omega_{\rm B}(k,\eta) = \frac{\rho_{\rm B}}{\rho} =\frac{1}{8\pi \rho a^4} \int d^{3} p\,\, B_{i}(|\vec{p} - \vec{k}|)\, B^{i}(p).
\label{OMB}
\end{eqnarray}
The appearance of convolutions in the expressions  of the magnetic energy density is a direct consequence 
of the absence of a uniform component of the magnetic field whose background contribution, as 
repeatedly stressed, is vanishing. 
Notice also  that $\rho$ appearing in Eq. (\ref{OMB})  is the energy density of the {\em dominant} 
component of the fluid, so, for instance, in the radiation-dominated epoch 
$\rho\equiv \rho_{\rm r}$, $a^4 \rho_{\rm r} \sim {\rm constant}$ and $\Omega_{B}\simeq {\rm constant}$. 

From Eq. (\ref{p0il}), using the same notation as in  Eq. (\ref{p00lex}), the following simplified equation can  be 
obtained for the momentum constraint
\begin{equation}
k^2 ( {\cal H} \phi + \psi') = \frac{3}{2} {\cal H}^2 \biggl[ \frac{4}{3} ( R_{\nu} \theta_{\nu} + ( 1 - R_{\nu} ) \theta_{\gamma}) + 
\frac{F_{\rm B}(k)}{4 \pi \sigma a^4 \rho} + \theta_{\rm b} \Omega_{\rm b} + \theta_{\rm c} \Omega_{\rm c} \biggr],
\label{p0ilex}
\end{equation}
where $F_{\rm B}(k,\eta)$ is the Fourier transform of the generalized Lorentz force, i.e. 
\begin{equation}
 \vec{\nabla} \cdot [ (\vec{\nabla} \times \vec{B}) \times \vec{B}] = \int d^{3} k  F_{\rm B}(k) e^{ i \vec{k} \cdot \vec{x}},
\end{equation}
where $ F_{\rm B}(k)$ is given by the following convolution  
\begin{equation}
 \int d^{3} p [ ( \vec{k}\cdot \vec{p}) B_{j}(p)B^{j}(|\vec{k} - \vec{p}|) - (k_{i} - p_{i})B_{j}(|\vec{k} - \vec{p}|) B_{j}(|\vec{k} - \vec{p}|)
B^{j}(p)]. 
\end{equation}
Notice that a consistent treatment of fully inhomogeneous magnetic fields implies that 
\begin{equation}
\Omega_{\rm B} \ll 1, ~~~~~~~~~\frac{F_{\rm B}}{4\pi k^2  \rho a^4} \ll 1,
\label{closure}
\end{equation}
in order not to over-close the Universe. 

In the ideal MHD limit\footnote{Notice that, so far, different quantities have been denoted by $\sigma$ (with different indices), namely the Thompson 
cross section (i.e. $\sigma_{\rm T}$), the conductivity (i.e. $\sigma$) and the fractional contribution of each species $\lambda$ to the 
anisotropic stress (i.e. $\sigma_{\lambda}$). Since all the quantities are properly defined and used, there should be no 
confusion. We felt that it would be even more confusing if we changed the usually accepted notations for these three important quantities.}, 
i.e. $\sigma\to \infty$, the contribution of $F_{\rm B}(k)$ disappears from Eq. (\ref{p0ilex}). This
occurrence is not general: it will be shown in a moment that, even in the ideal MHD limit, the generalized Lorentz force 
does contribute  both to the baryon evolution equation and to the anisotropic stress.
In fact the evolution equation for the baryons becomes, in Fourier space and with the conventions used so far
\begin{equation}
\theta_{\rm b}' + {\cal H} \theta_{\rm b} = k^2 \phi + \frac{F_{\rm B}(k)}{ 4\pi a^4 \rho_{\rm b}}.
\label{redbaryon}
\end{equation}
By taking the trace of  Eq. (\ref{pijl}) it is easy to obtain 
\begin{equation}
\psi'' + ( 2 \psi' + \phi') {\cal H} + (2 {\cal H}' + {\cal H}^2) \phi - \frac{k^2}{3} ( \phi - \psi) = \frac{1}{2} {\cal H}^2 [ ( R_{\nu} \delta_{\nu} + (1 - R_{\nu})) 
\delta_{\gamma} + \Omega_{\rm B}(k)],
\label{pijlex}
\end{equation}
where the contribution of the other species vanishes since both baryons and CDM particles have a vanishing barotropic index, i.e. $w_{\rm b} = w_{\rm c} =0$.

By taking now the difference between Eqs. (\ref{pijl}) and (\ref{pijlex}), we can get the equation for the anisotropic stress, i.e. for the perturbed components 
$i\neq j$ of the Einstein equations:
\begin{equation}
\partial_{i}\partial^{j}(\phi -\psi) - \frac{1}{3}\nabla^2 (\phi -\psi) = 8\pi G a^2 \rho \biggl\{ \Sigma_{i}^{j} + \frac{1}{4\pi\rho a^{4}} \biggl[ B_{i}B^{j} - 
\frac{1}{3} \vec{B}^2 \delta_{i}^{j}\biggr]\biggr\}.
\label{pineqjl}
\end{equation}
Applying now the differential operator $\partial_{j}\partial^{i}$ to Eq. (\ref{pineqjl}), the following relation can be obtained 
\begin{equation}
\nabla^4(\phi -\psi) = 12\pi G a^2 \biggl[ {\cal Q} + \frac{1}{4\pi a^4 \rho} ( \partial_{i} B_{j} \partial^{j}B^{i} - \frac{1}{3} \nabla^2 \vec{B}^2)\biggr]
\label{shex0}
\end{equation}
where, following the conventions of Eq. (\ref{shdef}),
\begin{equation}
{\cal Q} =\partial_{i}\partial^{j} \Sigma_{j}^{i} = - k^2 \sigma_{\nu}(p_{\nu} + \rho_{\nu}) = - \frac{4}{3} k^2 \rho_{\nu}.
\label{shex}
\end{equation}
In order to write Eq. (\ref{shex}) in an explicit form  
it has been assumed that the only anisotropic stress arising from the fluid sources 
is the one related to the quadrupole moment of the neutrino phase-space density.
Going to Fourier space, Eq. (\ref{shex0}) becomes 
\begin{equation}
k^2 (\phi - \psi) = \frac{9}{2}{\cal H}^2  \biggl[- \frac{4}{3} \sigma_{\nu}  - \sigma_{\rm B}(k)\biggr],
\label{pineqjex}
\end{equation}
having defined
\begin{equation}
\sigma_{\rm B}= \frac{\Omega_{\rm B}(k)}{3} - \frac{F_{\rm B}(k)}{4 \pi a^4 \rho k^2}.
\label{sigmaB}
\end{equation}
The quantity $\sigma_{\rm B}$ can be interpreted as the anisotropic stress arising thanks to the 
inhomogeneous magnetic field and,
in the force-free case, $\sigma_{\rm B} = \Omega_{\rm B}/3$.

In order to discuss the initial conditions for the CMB anisotropies, the usual procedure is to solve 
the system of equations for the fluctuations while  the Fourier modes of the 
various fields are outside the horizon, i.e. $k\eta < 1$. Different sorts  
of initial conditions are, in principle, possible. An interesting situation arises when 
both $\psi$ and $\phi$ are, in the first approximation, constant outside the horizon during the 
radiation-dominated epoch. In this case, and in the absence of magnetic fields, it is known that the system 
of scalar metric perturbations admits an {\em adiabatic} mode.
Adiabatic means, in the present context, that the total fluctuations in the entropy density 
of the CDM--baryon--radiation fluid vanish at large length  scales. This condition 
implies specific relations between the various density contrasts deep within the radiation-dominated epoch, i.e.
\begin{equation}
\delta_{\rm c} \simeq \delta_{\rm b} \simeq \frac{3}{4} \delta_{\nu} \simeq \frac{3}{4} \delta_{\gamma}.
\label{adiabaticity}
\end{equation}
In the following, after reviewing the specific analytic form of the adiabatic solution, 
new solutions, arising in the presence of an inhomogeneous magnetic field, will 
be illustrated. 

\subsection{The standard adiabatic scenario}

If the magnetic fields are absent, i.e. $F_{\rm B} = \Omega_{B} =0$ the standard system of equations is recovered
and the initial conditions for the various fields
 can be found by solving, simultaneously,  the baryon--photon system (i.e. Eqs. (\ref{baryon1})--(\ref{baryon2}) and Eqs. (\ref{phot1})--(\ref{phot2})), 
the CDM and neutrinos equations (i.e. Eqs. (\ref{CDM1})--(\ref{CDM2}) and Eqs. (\ref{nu1})--(\ref{nu3})) together 
with the perturbed Einstein equations whose specific form has been derived above from Eq. (\ref{p00lex}) to Eq. (\ref{pineqjex}).
In the following, only the final result will be given:
\begin{eqnarray}
&& \overline{\delta}_{\rm b} = 
\overline{\delta}_{\rm c} = - \frac{3}{2} \phi_{0} - \frac{\left( 525 + 188\,R_{\nu} + 16\,R_{\nu}^2 \right) }{60\,\left( 25 + 2\,R_{\nu} \right) }\phi_{0} k^2 \eta^2,
\label{ad1}\\
&& \overline{\delta}_{\gamma} = \overline{\delta}_{\nu} 
= - 2 \phi_{0} - \frac{\left( 525 + 188\,R_{\nu} + 16\,R_{\nu}^2 \right) }{45\,\left( 25 + 2\,R_{\nu} \right) }\phi_{0} k^2 \eta^2,
\label{solde1}\\
&& \overline{\phi} = \phi_0  -\frac{\left( 75\, + 14\,\,R_{\nu} - 8\,\,R_{\nu}^2 \right) }{90\,\left( 25 + 2\,R_{\nu} \right) }\phi_0 k^2 \eta^2,
\label{solphi1}\\
&& \overline{\psi} 
= \biggl( 1 + \frac{2}{5} R_{\nu} \biggr)\phi_{0}  -\frac{\left( 75 + 79\,R_{\nu} + 8\,R_{\nu}^2 \right) }{90\,\left( 25 + 2\,R_{\nu} \right) }\phi_{0} k^2 \eta^2,
\label{solpsi1}\\
&& \overline{\theta}_{\nu} = \frac{\phi_0}{2}k^2 \eta - \frac{\left( 65 + 16\,R_{\nu} \right) }{36\,\left( 25 + 2\,R_{\nu} \right) }\phi_0 k^4 \eta^3,
\label{solthnu1}\\
&& \overline{\theta}_{\rm b} = \frac{\phi_{0}}{2} k^2 \eta -\frac{\left( 75\, + 14\,\,R_{\nu} - 8\,\,R_{\nu}^2 \right) }{360\,\left( 25 + 2\,R_{\nu} \right) }\phi_0 k^4 \eta^3,
\label{solthb1}\\
&&\overline{\theta}_{\rm c} = \frac{\phi_{0}}{2} k^2 \eta -\frac{\left( 75\, + 14\,\,R_{\nu} - 8\,\,R_{\nu}^2 \right) }{360\,\left( 25 + 2\,R_{\nu} \right) }\phi_0 k^4 \eta^3,
\label{solthc1}\\
&& \overline{\theta}_{\gamma} =  \frac{\phi_0}{2}k^2 \eta -\frac{\left( 25 + 8\,R_{\nu} \right) }{20\,\left( 25 + 2\,R_{\nu} \right) }\phi_{0} k^2 \eta^2,
\label{solthgam1}\\
&& \overline{\sigma}_{\nu} = \frac{\phi_{0}}{15} k^2 \eta^2  - \frac{\left( 65 + 16\,R_{\nu} \right) }{540\,\left( 25 + 2\,R_{\nu} \right) }\phi_0 k^4 \eta^4,
\label{solsig1}
\end{eqnarray}
where the overline in the various quantities has been introduced for future notational convenience. 
Concerning this solution, a few comments are in order:
\begin{itemize}
\item{} while the leading order of the solution is usually quoted (see for instance \cite{MB,turok1}), the second 
is useful in the present case, in  order to check what happens at the horizon crossing 
of a given mode (i.e. $k\eta \sim  1$) and in order to compare with the situation where the magnetic field is present;
\item{} while to leading order all the peculiar velocities of the plasma coincide, to 
next-to-leading order they are different;
\item{}  the adiabaticity condition given in Eq. (\ref{adiabaticity}) holds not only to leading order but also to next-to-leading order in $|k\eta|^2$.
\end{itemize}

\subsection{Force-free case}

Consider now the case when the magnetic field is force-free, and suppose,
 as previously discussed, that the solution at large scale is, in the first approximation, a constant mode 
for the longitudinal fluctuations of the metric. Then, the solution can be parametrized as 
\begin{equation}
\phi \simeq \phi_{0} + C_{\phi} k^2 \eta^2,~~~~~~~~~~~~~\psi \simeq \psi_{0} + C_{\psi} k^2\eta^2.
\label{ans1}
\end{equation}
We are interested in the solution of the system deep in the radiation epoch where $a(\eta) \sim \eta$ and the dominant 
component of the fluid energy-momentum tensor consist of photons and neutrinos, i.e. $ \Omega_{\rm b} \ll 1$ and $ \Omega_{\rm c} \ll 1$.
From Eq. (\ref{p00lex}), the density contrasts for photons and neutrinos can be written as 
\begin{eqnarray}
&&\delta_{\gamma} = - 2 \phi_{0} - \Omega_{\rm B} + A_{\gamma} k^2 \eta^2,
\label{agamma}\\
&& \delta_{\nu} = - 2 \phi_0 - \Omega_{\rm B} + A_{\nu} k^2 \eta^2,
\label{anu1}
\end{eqnarray}
where the constants $A_{\gamma}$ and $A_{\nu}$ are related to $C_{\phi}$ and $C_{\psi}$ by the following algebraic 
equation
\begin{equation}
C_{\phi} + 2 C_{\psi} + \frac{\psi_{0}}{3} = - \frac{1}{2} [ R_{\nu} A_{\nu} + (1 - R_{\nu}) A_{\gamma}].
\label{C0}
\end{equation}
Inserting  Eqs. (\ref{ans1}) together with Eq. (\ref{agamma}) 
into Eqs. (\ref{phot1}) and (\ref{phot2}), the following parametrization for the peculiar 
velocity of the photons can be derived
\begin{equation}
\theta_{\gamma } = k^2 \biggl[ \frac{\phi_{0}}{2} - \frac{\Omega_{\rm B}}{4}\biggr] \eta + D_{\gamma }k^4 \eta^3
\label{dgamma}
\end{equation}
together with two algebraic relations determining the constants 
\begin{eqnarray}
&& 3 D_{\gamma} = \frac{A_{\gamma}}{4} + C_{\phi}, 
\label{C1}\\
&& A_{\gamma} = -\frac{\phi_{0}}{3} + \frac{\Omega_{\rm B}}{6} + 4 C_{\psi}.
\label{C2}
\end{eqnarray}
With a similar procedure  the baryon and CDM fluids can be analysed: from Eqs. (\ref{baryon1})--(\ref{baryon2}) 
and Eqs. (\ref{CDM1})--(\ref{CDM2}), the solution can be written as follows
\begin{eqnarray}
&& \theta_{\rm b} = \frac{k^2 \phi_{0}}{2} \phi_{0}\eta + D_{\rm b} k^4 \eta^3,
\label{thetab}\\
&&  \theta_{\rm c} = \frac{k^2 \phi_{0}}{2} \phi_{0}\eta + D_{\rm c} k^4 \eta^3,
\label{thetac}
\end{eqnarray}
and 
\begin{eqnarray}
&& \delta_{\rm b} = - \frac{3}{4}( 2 \phi_0 + \Omega_{{\rm B}}) + A_{\rm b} k^2 \eta^2,
\label{deltab}\\
&&  \delta_{\rm c} = - \frac{3}{4}( 2 \phi_0 + \Omega_{{\rm B}}) + A_{\rm c} k^2 \eta^2.
\label{deltac}
\end{eqnarray}
The constants appearing in Eqs. (\ref{thetab})--(\ref{deltac}) should satisfy
\begin{eqnarray}
&& A_{\rm b,c} = 3 C_{\psi} - \frac{\phi_{0}}{4},
\label{abc}\\
&& 4 D_{\rm b,c} = C_{\phi},
\label{dbc}
\end{eqnarray}
where the notation $A_{\rm b, c}$ and $D_{\rm b, c}$ means 
that the relations hold, separately, for the baryon and CDM integration 
constants.

The  
anisotropic stress $\sigma_{\nu}$  appears both  in Eqs. (\ref{nu2}) and (\ref{nu3}) and in Eq. (\ref{pineqjex}),
determining the difference between the two longitudinal fluctuations of the metric. 
Hence, from Eqs. (\ref{pineqjex}) and (\ref{nu1})--(\ref{nu2}) the neutrinos fluid variables are determined to be 
\begin{eqnarray}
&&\sigma_{\nu} = \frac{k^2 \eta^2 }{6 R_{\nu} } ( \psi_{0} - \phi_{0}) - \frac{\Omega_{\rm B}}{4 R_{\nu}} + \frac{D_{\nu}}{15} k^4 \eta^4,
\label{signu}\\
&& \theta_{\nu} = k^2 \biggl[ \frac{\phi_{0}}{2} - \frac{\Omega_{\rm B}}{4} \frac{R_{\nu} -1}{R_{\nu}}\biggr] + D_{\nu} k^4 \eta^3, 
\label{thetnu}\\
&& \delta_{\nu} = - ( 2 \phi_{0} + \Omega_{\rm B}) = A_{\nu} k^2 \eta^2,
\end{eqnarray}
subject to the constraints 
\begin{eqnarray}
&&3 D_{\nu} - \frac{A_{\nu}}{4}= C_{\phi} - \frac{\phi_{0}}{15} + \frac{\Omega_{\rm B}}{30} \frac{R_{\nu} - 1}{R_{\nu}},
\label{dnu}\\
&& A_{\nu} = - \frac{\phi_{0}}{3} + \frac{\Omega_{\rm B}}{6}  \frac{R_{\nu} - 1}{R_{\nu}} + 4 C_{\psi},
\label{anu}\\
&& C_{\phi} - C_{\psi} =  - \frac{2}{5} D_{\nu}.
\label{dnu2}
\end{eqnarray}
The lowest order of  Eq. (\ref{nu3}) also implies
\begin{equation}
\psi_{0} = \phi_{0} \biggl( 1 + \frac{2}{5} R_{\nu}\biggr)  - \frac{\Omega_{\rm B}}{5} (R_{\nu} -1).
\label{difference}
\end{equation}
 
Up to now all the evolution equations for the fluid variables have been 
consistently solved in terms of a set of unknown constants satisfying a set of (algebraic) consistency conditions.
From the remaining two Einstein equations (\ref{p0ilex}) and (\ref{pijlex}), two further constraints on the various 
constants can be obtained, namely
\begin{eqnarray}
&& C_{\phi} + 2 C_{\psi} = 2 D_{\nu} R_{\nu} + 2 ( 1 - R_{\nu}) D_{\gamma},
\label{0ieq}\\
&& 6 C_{\psi} + C_{\phi} = - \frac{2}{15} R_{\nu} \phi_{0} + \frac{\Omega_{\rm B}}{15} (R_{\nu} - 1) + \frac{1}{2} [ A_{\nu} R_{\nu} + ( 1 - R_{\nu}) A_{\gamma}]. 
\label{ijeq}
\end{eqnarray}
Since two of the algebraic equations are not independent, the system of ten algebraic equations in ten unknowns 
becomes
\begin{eqnarray}
&& A_{\rm b} = A_{\rm c}= -\biggl[\frac{69 - 61\,R_{\nu} - 8\,R_{\nu}^2}{60\,\left( 25 + 2\,R_{\nu} \right) }\Omega_{\rm B} +
\frac{525 + 188\,R_{\nu} + 16\,R_{\nu}^2}{60\,\left( 25 + 2\,R_{\nu} \right) }\phi_0\biggr] 
\nonumber\\
&& A_{\gamma} =  - \biggl[ -\frac{237 + 152\,R_{\nu} + 16\,R_{\nu}^2}{90\,\left( 25 + 2\,R_{\nu} \right) }  \Omega_{\rm B}   + 
 \frac{\left( 525 + 188\,R_{\nu} + 16\,R_{\nu}^2 \right) }{45\,\left( 25 + 2\,R_{\nu} \right) }\phi_{0}\biggr],
\nonumber\\
&& A_{\nu} = - \biggl[\frac{375 - 207\,R_{\nu} - 152\,R_{\nu}^2 - 16\,R_{\nu}^3}{90\,R_{\nu}\,\left( 25 + 2\,R_{\nu} \right) } \Omega_{\rm B} + 
 \frac{\left( 525 + 188\,R_{\nu} + 16\,R_{\nu}^2 \right) }{45\,\left( 25 + 2\,R_{\nu} \right) }\phi_{0}\biggr] ,
\nonumber\\
&& C_{\phi} =  -\biggl[ \frac{6 - 8\,R_{\nu} + 2\,R_{\nu}^2}{45\,\left( 25 + 2\,R_{\nu} \right) } \Omega_{\rm B} +  
\frac{75 + 14\,R_{\nu} - 8\,R_{\nu}^2}{90\,\left( 25 + 2\,R_{\nu} \right) } \phi_{0}\biggr],
\nonumber\\
&& C_{\psi} =  -\biggl[ \frac{69 - 61\,R_{\nu} - 8\,R_{\nu}^2}{180\,\left( 25 + 2\,R_{\nu} \right) }\Omega_{\rm B} 
+ \frac{\left( 75 + 79\,R_{\nu} + 8\,R_{\nu}^2 \right) }{90\,\left( 25 + 2\,R_{\nu} \right) }\phi_{0}\biggr],
\nonumber\\
&&D_{\gamma} = 
- \biggl[ \frac{\left( 25 + 8\,R_{\nu} \right) }{20\,\left( 25 + 2\,R_{\nu} \right) }\phi_{0} 
-\frac{\left( 7 + 8\,R_{\nu}  \right) }{40\,\left( 25 + 2\,R_{\nu} \right)}\Omega_{\rm B} \biggr] ,
\nonumber\\
&& D_{\nu} =       - \biggl[\frac{45 - 29\,R_{\nu} - 16\,R_{\nu}^2}{72\,R_{\nu}\,\left( 25 + 2\,R_{\nu} \right) }\Omega_{\rm B}
+\frac{\left( 65 + 16\,R_{\nu} \right) }{36\,\left( 25 + 2\,R_{\nu} \right) }\phi_0\biggr],
\nonumber\\
&& D_{\rm b} = D_{\rm c} = \frac{\phi_{0}}{2} k^2 \eta  -\biggl[ \frac{6 - 8\,R_{\nu} + 2\,R_{\nu}^2}{180\,\left( 25 + 2\,R_{\nu} \right) } \Omega_{\rm B} +  
\frac{75 + 14\,R_{\nu} - 8\,R_{\nu}^2}{360\,\left( 25 + 2\,R_{\nu} \right) } \phi_{0}\biggr].
\label{arbconst}
\end{eqnarray}

Recalling the form of the standard  solution reported in Eqs. (\ref{ad1})--(\ref{solsig1}), and using the results obtained so far, the full solution can be written as
\begin{eqnarray}
&&
\delta_{\rm b, c} = \overline{\delta}_{\rm b, c} - \frac{3}{4} \Omega_{\rm B} 
- \biggl[\frac{69 - 61\,R - 8\,R_{\nu}^2}{60\,\left( 25 + 2\,R_{\nu} \right) }\biggr]\Omega_{\rm B}k^2 \eta^2,
\label{dbOM}\\
&&\delta_{\gamma} = \overline{\delta}_{\gamma} - \Omega_{\rm B} 
+ \biggl[\frac{237 + 152\,R_{\nu} + 16\,R_{\nu}^2}{90\,\left( 25 + 2\,R_{\nu} \right) } \biggr] 
\Omega_{\rm B} k^2 \eta^2,
\label{dgOM}\\
&& \delta_{\nu} = \overline{\delta}_{\nu} -\Omega_{\rm B} 
- \biggl[\frac{375 - 207\,R_{\nu} - 152\,R_{\nu}^2 - 16\,R_{\nu}^3}{90\,R_{\nu}\,\left( 25 + 2\,R_{\nu} \right) } \biggr]
\Omega_{\rm B} k^2 \eta^2,
\label{dnOM}\\
&& \phi = \overline{\phi} 
-\biggl[\frac{6 - 8\,R_{\nu} + 2\,R_{\nu}^2}{45\,\left( 25 + 2\,R_{\nu} \right) } \Omega_{\rm B} \biggr]k^2 \eta^2,
\label{phOM}\\
&& \psi = \overline{\psi} 
-\biggl[ \frac{69 - 61\,R_{\nu} - 8\,R_{\nu}^2}{180\,\left( 25 + 2\,R_{\nu} \right) }\Omega_{\rm B}\biggr] k^2 \eta^2,
\label{psOM}\\
&&\theta_{\gamma} = \overline{\theta}_{\gamma} - \biggl[\frac{\Omega_{\rm B}}{4} k^2 \eta 
+  \frac{\left(  7 + 8\,R_{\nu} \right)   }{40\,\left( 25 + 2\,R \right)}\Omega_{\rm B} \biggr] k^4\eta^3,
\label{thgOM}\\
&&\theta_{\nu} =     \overline{\theta}_{\nu} 
- \frac{\Omega_{\rm B}}{4} \frac{R_{\gamma}}{R_{\nu}}  k^2 \eta     
- \biggl[\frac{45 - 29\,R_{\nu} - 16\,R_{\nu}^2}{72\,R_{\nu}\,\left( 25 + 2\,R_{\nu} \right) }\Omega_{\rm B}\biggr] k^4 \eta^3,
\label{thnOM}\\
&&\theta_{\rm b} =\theta_{\rm c} = \overline{\theta}_{\rm b}  -
 \biggl[\frac{6 - 8\,R_{\nu} + 2\,R_{\nu}^2}{180\,\left( 25 + 2\,R_{\nu} \right) } \Omega_{\rm B}\biggr]  k^4 \eta^3,
\label{thbOM}
\end{eqnarray}
where the overline in the various quantities in the left-hand side denote the adiabatic solution derived above.

Concerning these results,  the following comments  are in order:
\begin{itemize}
\item{} unlike the standard case, the difference between the two constant modes of 
the longitudinal fluctuations of the metric is also determined by the magnetic energy density; 
in the limit $\Omega_{\rm B} \to 0$, the standard expression (obtained in the previous subsection) is recovered;
\item{} the adiabaticity condition is enforced to lowest order in $k\eta$, but it is violated at the next-to-leading 
order;
\item{} the force-free approximation neglects, by assumption, the Lorentz force and this is the 
rationale for the equality, for instance, between $\theta_{\rm b}$ and $\theta_{\rm c}$: in the 
absence of Lorentz force the magnetic field does not impart gradients to the baryonic 
component of the fluid;
\item{} the presence of magnetic fields introduces a further source of anisotropic stress as it can be 
appreciated from the chain of (first-order) relations $ \psi - \phi \simeq \overline{\psi} - \overline{\phi} - \Omega_{\rm B} ( R_{\nu} - 1)/5$ 
which follows from Eq. (\ref{difference}).
\end{itemize}
Since the adiabaticity condition is only realized to lowest order in $k\eta$ it is legitimate to name the solution 
presented in Eqs. (\ref{dbOM})--(\ref{thbOM}) {\em quasi-adiabatic}. The corrections to adiabaticity are of order 
$\Omega_{\rm B} k^2 \eta^2$, so they are suppressed outside the horizon and also, by virtue of Eq. (\ref{closure}),  by $\Omega_{\rm B}$. 
Notice that three regimes emerge naturally. The quasi- adiabatic regime where 
$\Omega_{\rm B}(k) \leq \phi_{0}(k)$, the isocurvature regime $\Omega_{\rm B}(k) > \phi_{0}(k)$ and the fully 
adiabatic regime, i.e. $\Omega_{\rm B}(k) \ll \phi_{0}(k)$.

The force-free limit may be (approximately) realized in nature in some specific (but rather interesting) cases.
For instance, it could happen that magnetic fields are generated in a so-called maximally helical configuration. 
Recall, in fact, that given a magnetic field configuration, in a MHD description,
 it is always possible to  define the related magnetic helicity \cite{biskamp}, which is 
\begin{equation}
\int d^3 x \vec{A} \cdot \vec{B}.
\end{equation}
Now, at finite conductivity, the evolution of the magnetic helicity can be written as 
\begin{equation}
\frac{d}{d\eta} \int d^3 x \vec{A} \cdot \vec{B} = - \frac{1}{4\pi \sigma} \int d^{3} x \vec{B} \cdot \vec{\nabla}\times \vec{B},
\end{equation}
where the expression appearing in the right-hand side is called magnetic gyrotropy \cite{biskamp} and  measures the 
number of contact points in the magnetic field lines. Clearly, if the magnetic gyrotropy is 
maximal, the orthogonal combination, namely the Lorentz force $\vec{\nabla}\times \vec{B} \times \vec{B}$ , will be 
minimal or even zero. Maximally helical configurations can be very relevant in the development of MHD turbulence \cite{biskamp}
and have been extensively studied in a cosmological context.
(see also \cite{giovannini} for a general discussion and \cite{campanelli} for a recent paper). In 
particular it should be noticed that helical configurations can be produced at the electroweak time 
by coupling the hypercharge field to a pseudo-scalar \cite{gm1} (see also \cite{gm2}). 

\subsection{Contribution of the Lorentz force}

Consider now the case when the magnetic field  is not  force-free.
The major difference with  the force-free case is that the evolution equation
for the peculiar velocity of the baryons receives a contribution from the magnetic field gradient. Because of the tightly 
coupled dynamics of baryons and photons, also the photon peculiar velocity will be modified. Since photons are 
related to neutrinos by the Einstein equations, the whole solution will be modified.

Equation (\ref{baryon1}) can be written, after multiplying both sides by the scale factor $a(\eta)$, as 
\begin{equation}
(a \theta_{\rm b})' = k^2 \phi_{0}a + \frac{F_{\rm B}}{4\pi \rho_{\rm b} a^3},
\label{baryon3}
\end{equation}
where it has been assumed that $\phi$ has, to lowest order, a constant solution denoted, as usual, by $\phi_{0}$.
The second term in the right-hand side of Eq. (\ref{baryon3}) is then constant and  direct integration gives
\begin{equation}
\theta_{\rm b} = \frac{\phi_{0}}{2} k^2 \eta + \frac{F_{\rm B}}{4 \pi \rho_{\rm b} a^3}.
\label{baryon4}
\end{equation}
From Eq. (\ref{baryon2}), using Eq. (\ref{baryon4}), the density contrast can be determined  up to a constant, which is fixed 
from the adiabaticity condition:
\begin{equation}
\delta_{\rm b} = - \frac{3}{2} \phi_{0} - \frac{3}{4} \Omega_{\rm B} - \frac{ F_{\rm B}}{4\pi \rho_{\rm b} a^3} \eta.
\label{baryon5}
\end{equation}
The Hamiltonian constraint of Eq. (\ref{p00lex}) and the equation for the peculiar velocity of the photons imply 
\begin{eqnarray}
&& \delta_{\nu} \simeq \delta_{\gamma} \simeq - 2\phi_0 - \Omega_{\rm B},
\label{dens1}\\
&&  \theta_{\gamma} = \biggl( \frac{\phi_{0}}{2} - \frac{\Omega_{\rm B}}{4} \biggr) k^2 \eta.
\label{thgam5}
\end{eqnarray}
The CDM equations (\ref{CDM1}) and (\ref{CDM2}) will then turn out to be 
\begin{equation}
\delta_{\rm c} \simeq -2\phi_{0}, \,\,\,\,\, \theta_{\rm c} = \frac{\phi_{0}}{2} k^2 \eta.
\end{equation}
 The anisotropic stress of the neutrinos can be 
determined by solving Eq. (\ref{pineqjex}):
\begin{equation}
\sigma_{\nu} = \frac{k^2\eta^2}{6 R}( \psi_{0} - \phi_{0}) - \frac{3}{4} \frac{\sigma_{\rm B}}{R_{\nu}}.
\label{sigmanu}
\end{equation}
Inserting eq. (\ref{sigmanu}) into Eq. (\ref{nu2}) the peculiar velocity of the neutrino fluid can be determined to be  
\begin{equation}
\theta_{\nu} = \biggl( \frac{\phi_{0}}{2} - \frac{\Omega_{\rm B}}{4} + \frac{3}{4} \frac{\sigma_{\rm B}}{R_{\nu}}\biggr) k^2 \eta.
\end{equation}
Finally from Eq. (\ref{nu3}), using Eq. (\ref{sigmanu}), the difference between the two longitudinal fluctuations of the metric 
can be derived:
\begin{equation}
\psi_{0} = \phi_{0} \biggl( 1 + \frac{2}{5} R_{\nu}\biggr)  - \frac{R_{\nu}}{5} \Omega_{\rm B} + \frac{3}{5} \sigma_{\rm B}.
\end{equation}
The momentum constraint derived in Eq. (\ref{p0ilex})  is satisfied by virtue of the following identity 
\begin{equation}
\sigma_{\rm B} k^2 \eta - \frac{\Omega_{\rm B}}{3} + \Omega_{\rm b} \frac{F_{\rm B}}{4 \pi \rho_{\rm b} a^3}=0.
\end{equation}
Notice also that the contribution of the Lorentz force is suppressed, for $k\to 0$ with 
respect to $\Omega_{\rm B}$ since, roughly,
$F_{\rm B} \sim k^2 \Omega_{\rm B}$.

\renewcommand{\theequation}{4.\arabic{equation}}
\section{Synchronous gauge description}
\setcounter{equation}{0}
In the longitudinal description, the gauge freedom is completely specified and, hence, the so-called gauge modes are absent\footnote{According to the usual 
nomenclature, gauge modes are those modes arising when gauge freedom is not completely fixed, as in the synchronous gauge.}.
As a consequence \cite{brand}, 
in the longitudinal gauge the metric fluctuations correspond directly to a particular set of 
gauge-invariant combinations, namely the Bardeen potentials \cite{bardeen}. Having said this, 
it is important to appreciate that longitudinal and synchronous descriptions should be regarded as complementary and not as 
opposite. The synchronous description, for instance, is more convenient when modes based on anisotropic 
stresses are discussed, as in the case of the present analysis. Furthermore, the known (old) problem 
of spurious gauge modes is completely settled by now  since the work of Press and Vishniac \cite{PV1}. In some 
specific calculations it turns out to be useful to profit from the extra gauge freedom inherent in the synchronous
 description \label{gris}.
There are examples in the literature of isocurvature modes that are divergent, at early times, 
in the longitudinal gauge, but which are perfectly physical and regular in the 
synchronous coordinate system \cite{turok1}. Finally, it is important to consider 
the formulation of the problem of initial conditions in the synchronous gauge, since 
various numerical codes solving the Boltzmann hierarchy use, indeed, the synchronous description.

In the synchronous description the line element can be consistently perturbed in the form:
\begin{equation}
ds^2 = a^{2}(\eta) d\eta^2 - a^2(\eta) ( \delta_{ij} - h_{i j}) dx^{i} dx^{j},
\end{equation}
where the perturbed element of the metric is given by 
\begin{equation}
\delta g_{i j} = a^2 h_{ij}, \,\,\,\,\,\,\,\,\, \delta g^{i j} = - \frac{h^{ij}}{a^2}.
\label{pertsynch1}
\end{equation}
Sometimes the parametrization $\delta g_{ij} = 2 a^2 ( \psi_{\rm s} \delta_{i j} + \partial_{i} \partial_{j} E_{\rm s} ) $
is also employed \cite{brand}. The connection between the two parametrizations is immediate 
by separating, in Fourier space,  the trace of the  perturbation from its traceless part: 
\begin{equation}
h_{ij}(\eta,\vec{x}) = \int d^{3} k e^{i \vec{k} \cdot \vec{x}} \biggl[ \hat{k}_{i} \hat{k}_{j} h(k,\eta) + 6 \xi(k,\eta) \biggl(  
\hat{k}_{i} \hat{k}_{j} - \frac{1}{3} \delta_{ij}\biggr)\biggr],
\label{pertsynch2}
\end{equation}
where $\hat{k}^{i} = k^{i}/|\vec{k}|$.
Clearly, by performing an infinitesimal gauge transformation the $\delta g_{00}$ and $\delta g_{0i}$ parts of the metric (which are 
not perturbed in the synchronous parametrization) also transform. The gauge modes can be exactly identified by requiring 
that $ \delta g_{00} =0$ and $\delta g_{ij}=0$, for gauge transformations that preserve the synchronous nature 
of the coordinate system.  The first gauge mode corresponds to a spatial reparametrization of the constant-time hypersurfaces. 
As a consequence, in this mode the metric perturbation is constant and the matter density unperturbed.
The second gauge mode corresponds to a spatially dependent shift in the time direction. 

With the notation of Eqs. (\ref{pertsynch1}) and (\ref{pertsynch2}), the perturbed Einstein equations become 
\begin{eqnarray}
&& k^2 \xi - \frac{{\cal H}}{2} h' = \frac{3}{2} {\cal H}^2[ R_{\nu} \delta_{\nu} + ( 1 - R_{\nu}) \delta_{\gamma} + \Omega_{\rm B} 
+ \Omega_{\rm b} \delta_{\rm b} + \Omega_{\rm c} \delta_{\rm c}],
\label{00s}\\
&& k^2 \xi' = - \frac{3}{2} {\cal H}^2 \biggl[ \frac{F_{\rm B}}{4 \pi \sigma\rho} + \frac{4}{3} ( R_{\nu} \theta_{\nu} + 
( 1 - R_{\nu}) \theta_{\gamma}) + \Omega_{\rm b} \theta_{\rm b} + \Omega_{\rm c} \theta_{\rm c}\biggr],
\label{0is}\\
&& h'' + 2 {\cal H} h' - 2 k^2 \xi = 9 {\cal H}^2  \biggl[ \frac{1}{3} ( R_{\nu} \delta_{\nu} + ( 1 - R_{\nu}) \delta_{\gamma}) + 
\frac{\Omega_{\rm B}}{3}\biggr],
\label{ijs}\\
&& h'' + 6 \xi'' + 2 {\cal H} h' + 12 {\cal H} \xi' - 2 k^2 \xi = 9 {\cal H}^2  \biggl[\sigma_{\rm B} + \frac{4}{3} R_{\nu} \sigma_{\nu} \biggr],
\label{shs}
\end{eqnarray}
where the background equations (\ref{b1})--(\ref{b3}) have  already been used to eliminate the energy and pressure 
densities\footnote{In this section the density contrasts 
and the peculiar velocity field are named in the same way as in the longitudinal gauge. It is understood that the density 
contrasts and the peculiar velocity fields are {\em not} equal in the two gauges and are related by the transformations listed below (see Eqs. (\ref{T1}) and (\ref{T2})).}. 
By combining appropriately Eqs. (\ref{00s}) and (\ref{ijs}), it is possible to obtain a further useful equation
\begin{equation}
h'' + {\cal H} h' = 3 {\cal H}^2 [ 2 R_{\nu} \delta_{\nu} + 2 ( 1 - R_{\nu}) \delta_{\gamma} + 2 \Omega_{\rm B} + 
\Omega_{\rm b} \delta_{\rm b} + \Omega_{\rm c} \delta_{\rm c}].
\end{equation}
The evolution equations of the peculiar velocities and density contrasts of the various species of the 
plasma can be obtained by perturbing the covariant conservation of the energy-momentum tensor.
The result is the following:
\begin{eqnarray}
&& \delta_{\nu}' = - \frac{4}{3} \theta_{\nu} + \frac{2}{3} h',
\label{dnus}\\
&& \delta_{\gamma}' = - \frac{4}{3} \theta_{\gamma} + \frac{2}{3} h', 
\label{dgs}\\
&& \delta_{\rm b}' = - \theta_{\rm b} + \frac{1}{2} h' ,
\label{dbs}\\
&& \delta_{\rm c}' = - \theta_{\rm c} + \frac{h'}{2},
\label{dcs}\\
&& \theta_{\nu}' = -k^2 \sigma_{\nu} + \frac{k^2}{4} \delta_{\nu},
\label{thns}\\
&& \theta_{\gamma}' = \frac{k^2}{4} \delta_{\gamma},
\label{thgs}\\
&& \theta_{\rm b}' = - {\cal H} \theta_{\rm b}  + \frac{F_{\rm B}}{4 \pi \rho_{\rm b} a^4},
\label{thbs}\\
&& \theta_{\rm c}' = - {\cal H} \theta_{\rm c}
\label{thcs}\\
&& \sigma_{\nu}' = \frac{4}{15} \theta_{\nu} - \frac{3}{10} k {\cal F}_{\nu 3} + \frac{2}{15} h' + \frac{4}{5} \xi'.
\label{signus}
\end{eqnarray}
As anticipated the synchronous gauge modes can be made harmless by eliminating the constant solution 
for $h$ and by fixing, for instance, the CDM velocity field to  zero. These two requirements  specify 
the coordinate system completely.

The force-free solution can also be obtained within the synchronous description. In particular, for the metric fluctuations the solution is the following 
\begin{eqnarray}
&& \xi = - 2 C + \frac{R_{\nu} -1}{5} \Omega_{\rm B} + \biggl[ \frac{5 + 4 R_{\nu}}{6( 15 + 4 R_{\nu})} C - \frac{R_{\nu} -1}{60} \Omega_{\rm B} \biggr]k^2 \eta^2,
\label{xisol1}\\
&& h = \biggl[ \frac{\Omega_{\rm B}}{10} - C \biggr] k^2 \eta^2.
\label{hsol1}
\end{eqnarray}
The constant 
introduced in Eqs. (\ref{xisol1}) and (\ref{hsol1}) has been defined in order 
to match the standard notation usually employed in the literature (see for instance \cite{MB}) 
to characterize the adiabatic (inflationary) mode.
By solving Eqs. (\ref{dnus})--(\ref{signus}) the solution for the 
density contrasts 
\begin{eqnarray}
&& \delta_{\gamma} = -\Omega_{\rm B} + \biggl[ \frac{ 2 R_{\nu} + 3}{30} \Omega_{\rm B} - \frac{2}{3} C \biggr] k^2 \eta^2,
\label{dgamsols1}\\
&& \delta_{\nu} = - \Omega_{\rm B} + \biggl[ \frac{2 R_{\nu}^2 + 3 R_{\nu} - 5}{30 R_{\nu}} \Omega_{\rm B} - \frac{2}{3} C\biggr] k^2 \eta^2 ,
\label{dnusols1}\\
&& \delta_{\rm b} = - \frac{3}{4} \Omega_{\rm B} + \biggl[ \frac{R_{\nu} -1 }{20} \Omega_{\rm B} - \frac{C}{2} \biggr] k^2 \eta^2,
\label{dbsols1}\\
&& \delta_{\rm c} =   - \frac{3}{4} \Omega_{\rm B} + \biggl[ \frac{R_{\nu} -1 }{20} \Omega_{\rm B} - \frac{C}{2} \biggr] k^2 \eta^2,
\label{dcsols1}
\end{eqnarray}
and for the peculiar velocities 
\begin{eqnarray}
&& \theta_{\gamma} = - \frac{\Omega_{\rm B}}{4} k^2 \eta + \biggl[ \frac{2 R_{\nu} + 3 }{360} \Omega_{\rm B} - \frac{C}{18} \biggr],
\label{thgsols1}\\
&& \theta_{\nu} = - \frac{\Omega_{\rm B}}{4} k^2 \eta + \biggl[ \frac{ 2 R_{\nu}^2 + 7 R_{\nu} -9}{360 R_{\nu}} \Omega_{\rm B} - \frac{23 + 4 R_{\nu}}{15 + 4 R_{\nu}} \biggr],
\label{thnusols1}
\end{eqnarray}
can be obtained.
The solution derived in the previous equations of the 
present section can be transformed into the longitudinal gauge by using the appropriate transformation, i. e.
\begin{eqnarray}
&&  \phi_{\rm L} = - \frac{1}{2 k^2} [ ( 6 \xi + h)'' + {\cal H} ( 6\xi + h)'],
\nonumber\\
&& \psi_{\rm L} = - \xi + \frac{{\cal H}}{2 k^2} ( 6 \xi' + h'),
\nonumber\\
&& \delta^{(\lambda)}_{\rm L} = \delta^{(\lambda)}_{\rm s} + 3 ( w_{\lambda} + 1) \frac{{\cal H}}{2 k^2} ( h' + 6 \xi'), 
\nonumber\\
&& \theta^{(\lambda)}_{\rm L} = \theta^{(\lambda)}_{\rm s} - \frac{1}{2} ( h' + 6 \xi'), 
\label{T1}
\end{eqnarray}
where the subscripts refer to the quantities evaluated either in the longitudinal or in the synchronous gauge.
By setting $\Omega_{\rm B}=0$ in Eqs. (\ref{xisol1})--(\ref{hsol1}) and (\ref{dgamsols1}) --(\ref{thnusols1}),
we recover the standard adiabatic mode discussed in the appendix with the constant $C$ determined as  
\begin{equation}
C = \frac{15 + 4 R_{\nu}}{20} \phi_{0},
\end{equation}
where $\phi_{0}$ is the value of the adiabatic mode in the longitudinal gauge discussed previously.
Of course, the solutions of the longitudinal gauge can also be directly transformed into the synchronous gauge 
using the appropriate transformation, which can be easily derived:
\begin{eqnarray}
&& \xi = - \psi_{L} - \frac{{\cal H}}{a} \int a \phi_{L} d\eta,
\nonumber\\
&& h = 6 \psi_{L} + 6 \frac{{\cal H}}{a} \int a \phi_{L} d\eta, 
- 2 k^2 \int \frac{d\eta''}{a(\eta'')} \int^{\eta''} a(\eta') \phi_{L}(\eta') d\eta', 
\nonumber\\
&& \delta^{(\lambda)}_{s} = \delta^{(\lambda)}_{L} + 3 {\cal H} ( w + 1)\frac{1}{a} \int \phi_{L} a d\eta,
\nonumber\\
&& \theta^{(\lambda)}_{s} = \theta^{(\lambda)}_{L} - \frac{k^2}{a} \int a \phi_{L} d\eta.
\label{T2}
\end{eqnarray}

Up to now we always studied adiabatic (or quasi-adiabatic) solutions. However, also isocurvature solutions 
can be generalized to include fully inhomogeneous magnetic fields. In order to look for isocurvature 
solutions it is useful to recall that in looking for perturbative solutions (both in the longitudinal and in the synchronous gauge) 
there are various small parameters. One is certainly $k\eta$, which is small outside the horizon. However, \
also $\Omega_{\rm b}$ and $\Omega_{\rm c}$ are small parameters deep within the 
radiation-dominated epoch. Let us make this statement more precise by considering the 
scale factor 
\begin{equation}
a(\eta) = \biggl[ \biggl(\frac{\eta}{\eta_{ 1}}\biggl) + \biggl(\frac{\eta}{\eta_{1}}\biggl)^2 \biggr],
\label{interpolation}
\end{equation}
interpolating between the radiation-dominated phase for $\eta \ll \eta_{1}$ and the matter dominated  epoch for $ \eta \gg \eta_{1}$.
In this case we can also write 
\begin{eqnarray}
&&\Omega_{\rm b,c} = \overline{\Omega}_{\rm b,c} \frac{a(\eta) }{a(\eta) + 1},
\label{ombar1}\\
&& \Omega_{\rm B} =  \overline{\Omega}_{\rm B}\frac{1}{a(\eta) + 1},
\label{omBar2}
\end{eqnarray}
where the subscripts in Eq. (\ref{ombar1}) refer either to baryons or to CDM and where we took, for simplicity, 
$\eta_{1} =1$.

We can then insert Eqs. (\ref{interpolation}) and (\ref{ombar1})--(\ref{omBar2}) into the evolution equations 
for the perturbations in the synchronous gauge, regarding $\Omega_{\rm b, c}$ as small paramters
deep within the radiation-dominated epoch ($\eta\to 0$). The result of this procedure is summarized by the 
following solutions, which are valid in the case $\sigma_{\rm B} = \Omega_{\rm B}/3$:
\begin{eqnarray}
&& h \simeq ( - 4 \overline{\Omega}_{\rm b} \eta + 6 \overline{\Omega}_{\rm b} \eta^2) ,
\label{isoh}\\
&& \xi \simeq \frac{2}{3} \overline{\Omega}_{\rm b} \eta - \overline{\Omega}_{\rm b} \eta^2, 
\label{isoxi}
\end{eqnarray}
for the metric perturbations and 
\begin{eqnarray}
&& \delta_{\gamma} \simeq \biggl( - \frac{8}{3} \overline{\Omega}_{\rm b} \eta + 4 \overline{\Omega}_{\rm b} \eta^2 \biggr),
- \overline{\Omega}_{\rm B}( 1 - \eta + \eta^3),
\label{isodg}\\
&& \delta_{\rm b} = ( 1 - 2 \overline{\Omega}_{\rm b} + 3 \overline{\Omega}_{\rm b} \eta^2),
\label{isodb}\\
&& \delta_{\nu} \simeq \biggl( - \frac{8}{3} \overline{\Omega}_{\rm b} \eta + 4 \overline{\Omega}_{\rm b} \eta^2 \biggr),
\label{isodnu}\\
&& \delta_{\rm c} \simeq  2 \overline{\Omega}_{\rm b} + 3 \overline{\Omega}_{\rm b} \eta^2,
\label{isodc}\\
&& \theta_{\gamma} \simeq -\frac{1}{3} \overline{\Omega}_{\rm b} k^2\eta^2 - \frac{k^2}{16} \overline{\Omega}_{\rm B}( 4 \eta - 2 \eta^2 + \eta^4),
\label{isog}\\
&& \theta_{\nu} \simeq -\frac{1}{3} \overline{\Omega}_{\rm b} k^2\eta^2 - \frac{k^2}{16} \overline{\Omega}_{\rm B}( 4 \eta - 2 \eta^2 + \eta^4)\frac{R_{\gamma}}{R_{\nu}},
\label{isonu}\\
&& \theta_{\rm c} =0,
\label{isoc}\\
&& \sigma_{\nu} \simeq - \frac{2}{3} \frac{\overline{\Omega}_{\rm b}}{2 R_{\nu} + 15} k^2 \eta^3 
- \frac{\overline{\Omega}_{\rm B}}{4 R_{\nu}}( 1 - \eta + \eta^3), 
\label{isosigma}
\end{eqnarray}
for the fluid quantities. Clearly, for this mode, the adiabaticity condition is not satisfied. Furthermore, by transforming 
the solution to the Newtonian gauge, it is easy to check that the longitudinal 
fluctuations of the metric vanish for $\eta \to 0$. In the limit $\Omega_{\rm B}\to 0$ this mode 
reduces to the baryon isocurvature mode already discussed in \cite{turok1}. A similar solution can 
be obtained by changing $ \overline{\Omega}_{\rm b} \to \overline{\Omega}_{\rm c}$. 

\renewcommand{\theequation}{5.\arabic{equation}}
\section{Tight-coupling expansion in magnetoactive plasmas}
\setcounter{equation}{0}
One of the analytical tools often employed in the theory of 
the CMB anisotropies is the so-called tight-coupling 
expansion \cite{peebles}. Defining the differential optical depth as done in Eq. (\ref{diffopdep})
the {\em exact} tight-coupling limit is realized when 
$\sigma_{\rm T} \to \infty$ and $1/\tau' \to 0$. 
If tight coupling is {\em exact},
 photons and baryons 
are synchronized so well that the photon phase-space distribution 
is isotropic in the baryon rest frame. Since the photon distribution is 
isotropic, the resulting radiation is not polarized. However, as the 
decoupling time approaches there is a regime where $|1/\tau'| < 1$ without 
being zero. The idea is then to tailor a systematic expansion in powers 
of $|1/\tau'|$ and to consider not only the zeroth order but also 
higher orders depending upon the accuracy required by the problem. 
 The evolution equations to be discussed 
are essentially the radiative transfer equations which are, in turn, the 
Boltzmann equations written in terms of the brightness perturbations. 
Within the notation employed in the present paper, these equations are 
derived in the equations (\ref{DI})--(\ref{DU}) of the appendix. 
The brightness perturbations should be studied for each 
of the four Stokes parameters. The brightness perturbations  $\Delta_{\rm I}$,
connected with the first Stokes parameter I, are 
the temperature fluctuations, i.e. fluctuations in the intensity 
of the radiation field. As discussed in the appendix, recalling that $\hat{n}$ is the direction 
of the comoving-three momentum of the photon and defining 
$\mu = \hat{k}\cdot\hat{n}$, the evolution of the brightness perturbation is
\begin{equation}
\Delta_{\rm I}' + i k\mu \Delta_{\rm I} = ( \psi' - i k \mu \phi) + \tau'\biggl[ - \Delta_{\rm I} + \Delta_{{\rm I}0} +  \mu v_{\rm b} - 
\frac{1}{2} P_{2}(\mu) S_{0}\biggr],
\label{DI1}
\end{equation}
where, in our notation, $ v_{\rm b} = \theta_{\rm b}/(i k)$ and 
\begin{equation}
S_{0} = \Delta_{{\rm I}2} + \Delta_{{\rm Q}0} + \Delta_{{\rm Q}2},
\label{s0}
\end{equation}
where, with obvious notation (discussed in detail in the appendix) 
\begin{eqnarray}
&& \Delta_{\rm I}(\vec{k}, \hat{n}, \eta) = \sum_{\ell} (- i)^{\ell} (2 \ell + 1)\Delta_{{\rm I}\ell}(\vec{k},\eta) P_{\ell}(\mu), 
\nonumber\\
&& \Delta_{\rm Q}(\vec{k}, \hat{n}, \eta) = \sum_{\ell} (- i)^{\ell} (2 \ell + 1)\Delta_{{\rm Q}\ell}(\vec{k},\eta) P_{\ell}(\mu), 
\label{expD}
\end{eqnarray}
$\Delta_{{\rm I}\ell}$ and $  \Delta_{{\rm Q}\ell}$ being the  $\ell$-th multipole of the brightness 
function $\Delta_{{\rm I}}$ and $\Delta_{{\rm Q}}$.
 
The function $P_{2}(\mu) = (3 \mu^2 -1)/2$ in Eq. (\ref{s0}) is the Legendre 
polynomial of second order, which appears in the collision operator of the Boltzmann 
equation for the photons 
due to the directional nature of Thompson scattering.
The evolution equations for the brightness perturbations 
connected with the Q and U Stokes parameters are 
\begin{eqnarray}
&& \Delta_{\rm Q}' + i k\mu \Delta_{\rm Q} = \tau'[ - \Delta_{\rm Q} + \frac{1}{2}( 1 - P_{2}(\mu))S_0],
\label{DQ1}\\
&& \Delta_{\rm U}' + i k\mu \Delta_{\rm U} = \tau' - \Delta_{\rm U}.
\label{DU1}
\end{eqnarray}
Finally, in order to close the system, the baryon evolution equation 
should be added using the definition $v_{\rm b} = \theta_{\rm b}/(i k)$ either 
in Eq. (\ref{baryon1}) or in (\ref{redbaryon}).:
\begin{equation}
v_{\rm b}' + {\cal H} v_{\rm b} = - i k\phi - \frac{\tau'}{\alpha} [3i \Delta_{{\rm I}1} + v_{\rm b}] - i W_{B}(k),
\label{vb0}
\end{equation}
where 
\begin{eqnarray}
&&{\cal W}_{B}(k) =  \frac{F_{\rm B}}{4\pi k \rho_{\rm b} a^4 },
\label{W}\\
&& \alpha = \frac{3}{4} \frac{\rho_{\rm b}}{\rho_{\rm \gamma}}.
\label{alpha}
\end{eqnarray}
In (\ref{vb0}) the dipole of the brightness function appears directly. 
In fact,  using  the results derived in the appendix, the following 
chain of identities holds:
\begin{equation}
\theta_{\gamma} = \frac{3}{4} k {\cal F}_{\gamma 1} = 3 \Delta_{{\rm I}1}.
\end{equation}
where ${\cal F}_{\gamma 1}$  is the dipole of the photon phase-space 
distribution and $\Delta_{{\rm I}1}$ is the monopole of the 
brightness perturbation.

The idea is now to expand Eqs. (\ref{DI1}) and (\ref{DQ1}) 
in powers of the small parameter $ \epsilon = |1/\tau'|$, i.e. 
the inverse of the differential optical depth of Thompson 
scattering.  Before doing the expansion, it is useful to derive the hierarchy for the brightness 
functions in full analogy with what is  discussed in the appendix for the case of the neutrino phase-space 
distribution. 
To this aim, each side of Eqs. (\ref{DI1})--(\ref{DQ1}) and (\ref{vb0}) will be multiplied 
by the various Legendre polynomials  and the  integration  over $\mu$ will be performed.
Noticing that, from the orthonormality relation for Legendre polynomials (Eq. (\ref{norm}) of the appendix),
\begin{equation}
\int_{-1}^{1} P_{\ell}(\mu) \Delta_{\rm I} d\mu = 2 (-i)^{\ell} \Delta_{{\rm I} \ell},
\end{equation}
and recalling that
\begin{equation}
P_{0}(\mu) =1,\,\,\,\,\,P_{1}(\mu) = \mu,\,\,\,\,\,\,\,\,P_{2}(\mu)= \frac{1}{2}(3 \mu^2 -1),\,\,\,\,\,\, P_{3}(\mu) = \frac{1}{2}( 5\mu^3 -3 \mu),
\end{equation}
Eqs. (\ref{DI1})--(\ref{vb0}) allow the determination of the first three sets of equations for  the hierarchy of the brightness.
More specifically, multiplying Eqs. (\ref{DI1})--(\ref{DQ1}) and (\ref{vb0}) by $P_{0}(\mu)$ and integrating over $\mu$, the following relations can be obtained
\begin{eqnarray}
&& \Delta_{{\rm I}0}' + k \Delta_{{\rm I}1} =  \psi',
\label{L01}\\
&& \Delta_{{\rm Q}0} '+k \Delta_{{\rm Q}1} = \frac{\tau'}{2} [ \Delta_{{\rm Q}2} + \Delta_{{\rm I}2} - \Delta_{{\rm Q}0} ],
\label{L02}\\
&& v_{b}' + {\cal H} v_{\rm b} = - i k \phi - \frac{\tau'}{\alpha} ( 3 i \Delta_{{\rm I}1} + v_{b} ) - \frac{i}{a} W_{B}(k).
\label{L03}
\end{eqnarray}
If Eqs. (\ref{DI1})--(\ref{DQ1}) and (\ref{vb0}) are multiplied by  $P_{1}(\mu)$, the integration  over $\mu$ of the various terms implies 
\begin{eqnarray}
&& - \Delta_{{\rm I} 1}' - \frac{2}{3}k \Delta_{{\rm I}2} + \frac{k}{3} \Delta_{{\rm I}0} = - \frac{k}{3}  \phi + \tau' \biggl[ \Delta_{{\rm I} 1} + 
\frac{1}{3 i} v_{\rm b}\biggr],
\label{L11}\\
&& - \Delta_{{\rm Q}1}' - \frac{2}{3} k \Delta_{{\rm Q}2} + \frac{k}{3} \Delta_{{\rm Q}0} = \tau' \Delta_{{\rm Q} 1} 
\label{L12}\\
&& v_{b}' + {\cal H} v_{b} = - i k \phi - \frac{\tau'}{\alpha} ( 3 i \Delta_{{\rm I}1} + v_{b} )- \frac{i}{a} W_{B}(k).
\label{L13}
\end{eqnarray}
The same  procedure, using $P_{2}(\mu)$, leads to
\begin{eqnarray}
&& - \Delta_{{\rm I} 2}' - \frac{3}{5} k \Delta_{{\rm I}3} + \frac{2}{5} k \Delta_{{\rm I} 1} = \tau'\biggl[ \frac{9}{10} \Delta_{{\rm I}2} - \frac{1}{10} (\Delta_{{\rm Q}0} + 
\Delta_{{\rm Q} 2} )\biggr],
\label{L21}\\
&&  - \Delta_{{\rm Q} 2}' - \frac{3}{5} k \Delta_{{\rm Q}3} + \frac{2}{5} k \Delta_{{\rm Q} 1} = \tau'\biggl[ \frac{9}{10} \Delta_{{\rm Q}2} - \frac{1}{10} (\Delta_{{\rm Q}0} + 
\Delta_{{\rm I} 2} )\biggr],
\label{L22}\\
&& v_{b}' + {\cal H} v_{b} = - i k \phi - \frac{\tau'}{\alpha} \biggl( 3 i \Delta_{{\rm I}1} + v_{b} \biggr)- \frac{i}{a} W_{B}(k).
\end{eqnarray}
For $\ell\geq 3$ the hierarchy of the brightness can be determined in general terms by using the recurrence relation for the 
Legendre polynomials reported in Eq. (\ref{rec1}) of the appendix:
\begin{eqnarray}
&&\Delta_{{\rm I}\ell}'+ \tau' \Delta_{{\rm I}\ell} 
= \frac{k}{2 \ell + 1} [ \ell \Delta_{{\rm I}(\ell-1)} - (\ell + 1) \Delta_{{\rm I}(\ell + 1)}],
\nonumber\\
&& \Delta_{{\rm Q}\ell}'+ \tau' \Delta_{{\rm Q}\ell} 
= \frac{k}{2 \ell + 1} [ \ell \Delta_{{\rm Q}(\ell-1)} - (\ell + 1) \Delta_{{\rm Q}(\ell + 1)}].
\end{eqnarray}

We are now ready to compute the evolution of the various terms to a given order in the tight-coupling expansion parameter $\epsilon = |1/\tau'|$. 
After expanding the various moments of the brightness function and the velocity field in $\epsilon$ 
\begin{eqnarray}
&&\Delta_{{\rm I}\ell} = \overline{\Delta}_{{\rm I}\ell} + \epsilon \delta_{{\rm I}\ell},
\nonumber\\
&& \Delta_{{\rm Q}\ell} = \overline{\Delta}_{{\rm Q}\ell} + \epsilon \delta_{{\rm Q}\ell},
\nonumber\\
&&v_{\rm b} = \overline{v}_{\rm b} + \epsilon \delta v_{\rm b},
\end{eqnarray}
the obtained expressions can be inserted  into Eqs. (\ref{L01})--(\ref{L13}) and the evolution of the various moments of the brightness 
function can be found order by order.

To zeroth order in the tight-coupling approximation, the evolution equation for the baryon velocity field, i.e. Eq. (\ref{L03}), leads to:
\begin{equation}
\overline{v}_{b} = - 3 i  \overline{\Delta}_{{\rm I}1},
\label{vb1}
\end{equation} 
while Eqs. (\ref{L02}) and (\ref{L12}) lead, respectively, to
\begin{eqnarray}
&&\overline{\Delta}_{{\rm Q}0} = \overline{\Delta}_{{\rm I}2} + \overline{\Delta}_{{\rm Q}2}.
\nonumber\\
&& \overline{\Delta}_{{\rm Q}1} =0.
\label{int1}
\end{eqnarray}
Finally Eqs. (\ref{L21}) and (\ref{L22}) imply
\begin{eqnarray}
&& 9\overline{\Delta}_{{\rm I}2}  = \overline{\Delta}_{{\rm Q}0} + \overline{\Delta}_{{\rm Q}2},
\nonumber\\
&&  9\overline{\Delta}_{{\rm Q}2}  = \overline{\Delta}_{{\rm Q}0} + \overline{\Delta}_{{\rm I}2}.
\label{int2}
\end{eqnarray}
Taking together the four conditions expressed by Eqs. (\ref{int1}) and (\ref{int2}) we have, to zeroth order in the 
tight-coupling approximation:
\begin{eqnarray}
&&\overline{\Delta}_{{\rm Q}\ell} =0,\,\,\,\,\,\,\,\,\,\,\ell\geq 0,
\nonumber\\
&& \overline{\Delta}_{{\rm I}\ell} =0,\,\,\,\,\,\, \ell \geq 2.
\label{int3}
\end{eqnarray}
Hence, to zeroth order in the tight coupling, the relevant equations are 
\begin{eqnarray}
&& \overline{v}_{b} = - 3i \overline{\Delta}_{{\rm I}1},
\label{zerothorder1}\\
&& \overline{\Delta}_{{\rm I}0}' + k \overline{\Delta}_{{\rm I}1} =  \psi'
\label{zerothorder2}
\end{eqnarray}
This means, as anticipated, that to zeroth order in the tight-coupling expansion the CMB is not polarized since the quadrupole moment 
of the brightness is vanishing.
Summing up 
Eqs. (\ref{L11}) and (\ref{L13}) and using 
 Eq. (\ref{zerothorder1}) in order to eliminate $\overline{v}_{\rm b}$ from the obtained expression,
we get to the following equation
\begin{equation}
(\alpha + 1) \overline{\Delta}_{{\rm I}1}' + {\cal H} \alpha \overline{\Delta}_{{\rm I} 1} - \frac{k}{3} \overline{\Delta}_{{\rm I}0} 
= \frac{k}{3}  (\alpha+1) \phi + \frac{i \alpha}{3} W_{B}(k) .
\label{int4}
\end{equation}
Finally, the dipole term can be eliminated from Eq. (\ref{int4}) using Eq. (\ref{zerothorder2}). By doing so, Eq. (\ref{int4}) leads to  the wanted 
decoupled equation for the monopole:
\begin{equation}
\overline{\Delta}_{{\rm I}0}''  + \frac{\alpha'}{\alpha + 1} \overline{\Delta}_{{\rm I}0}' + \frac{k^2}{3 (\alpha +1)} \overline{\Delta}_{{\rm I}0} = 
 \biggl[ \psi'' + \frac{\alpha'}{\alpha +1} \psi' - \frac{k^2}{3} \phi \biggr] - \frac{\alpha}{\alpha + 1} \frac{ F_{\rm B}}{12\pi \rho_{\rm b} a^4}.
\label{monopole}
\end{equation}
In the limit $F_{\rm B} \to 0$ the usual decoupled equation for the evolution of the monopole is recovered.

The presence of the magnetic field modifies the evolution of the monopole. Hence, also the dipole will be modified 
since the relation (\ref{zerothorder2}) stipulates that the monopole is a source of the dipole.

To first order in the tight-coupling limit, the relevant equations can be obtained by keeping all terms of order $\epsilon$ and by using the 
first-order relations to simplify the expressions. From Eq. (\ref{L12}) the condition $\delta_{{\rm Q}1} =0$ can be derived; 
 from Eqs. (\ref{L02}) and  (\ref{L21})--(\ref{L22}), the following remaining conditions are obtained respectively:
\begin{eqnarray}
&& - \delta_{{\rm Q}0} + \delta_{{\rm I} 2} + \delta_{{\rm Q} 2} =0,
\label{firstord1}\\
&& \frac{9}{10} \delta_{{\rm I}2} - \frac{1}{10} [ \delta_{{\rm Q}0} + \delta_{{\rm Q}2} ]= \frac{2}{5} k \overline{\Delta}_{{\rm I}1},
\label{firstord2}\\
&&  \frac{9}{10} \delta_{{\rm Q}2} - \frac{1}{10} [ \delta_{{\rm Q}0} + \delta_{{\rm I}2} ]=0.
\label{firstord3}
\end{eqnarray}
Finally from Eqs. (\ref{firstord1})--(\ref{firstord3}) the remaining relations are: 
\begin{eqnarray}
&& \delta_{{\rm Q}0} = \frac{5}{4} \delta_{{\rm I}2},
\label{Cond1}\\
&& \delta_{{\rm Q}2} = \frac{1}{4} \delta_{{\rm I}2},
\label{Cond2}\\
&& \delta_{{\rm I}2} = \frac{8}{15} \overline{\Delta}_{{\rm I}1}.
\label{Cond3}
\end{eqnarray}
Condition (\ref{Cond3}) can be also written 
\begin{equation}
\Delta_{{\rm I} 2} = \epsilon \delta_{{\rm I}2}= \frac{8}{15} \frac{k}{\tau'} \overline{\Delta}_{{\rm I}1}.
\label{Cond4}
\end{equation}
Now, from Eqs. (\ref{Cond1}) and (\ref{Cond2}), the quadrupole moment of $\Delta_{\rm Q}$ is proportional to the quadrupole of $\Delta_{\rm I}$, which is, in turn, 
proportional to the dipole evaluated to first order in $\epsilon$. But $\Delta_{\rm Q}$ measures exactly 
the degree of  linear polarization of the radiation field. So, to first order in the tight-coupling expansion, the CMB {\em is} linearly 
polarized. Furthermore, as discussed in Eq. (\ref{monopole}) the presence of the magnetic field modifies the evolution of both the 
monopole and  the dipole. Thus, according to Eq. (\ref{Cond4}), also the polarization will be modified.

All the the solutions discussed in the previous sections can be used to determine the initial conditions for the 
evolution of the brightness using the tight-coupling expansion. Consider, as a possible example, the case when initial 
conditions are set deep within the radiation epoch. In this case, recalling Eq. (\ref{alpha}), Eq. (\ref{monopole}) can 
be written for, $\alpha \ll 1$, as 
\begin{equation}
\Delta_{0}'' + \omega^2 \Delta_{0} = - \omega^2 ( \phi + \psi) - \frac{F_{\rm B}}{16\pi \overline{\rho}_{\gamma}},
\label{monopole2}
\end{equation}
where $ \Delta_{0} = \overline{\Delta}_{{\rm I}0} - \psi$ and $ \omega = k/\sqrt{3}$; we also posit $a^4 \rho_{\gamma} = \overline{\rho}_{\gamma}$.
The solution of Eq. (\ref{monopole2}) can be easily obtained in terms of arbitrary integration constants.
These constants are fixed by specifying the initial conditions for the velocity field $v_{\rm b}$, which determines, according 
to Eq. (\ref{vb1}), the value of the dipole.
Suppose, for instance, to be interested in the case of quasi-adiabatic initial conditions in the presence of the reduced Lorentz force.
Then, from Eqs. (\ref{baryon4}) and (\ref{vb1}), recalling that $\theta_{\rm b} = i k v_{b}$, 
\begin{equation}
\overline{\Delta}_{{\rm I}1} = \frac{k\phi_{0}}{6} \eta + \frac{F_{\rm B}}{12 \pi \rho_{\rm b}a^3}.
\end{equation}
From Eq. (\ref{L13}) the initial condition for $\Delta_{0}$ can be determined to be:
\begin{equation}
\Delta_{0} \simeq - \frac{3}{2}\biggl( 1 + \frac{4}{15} R_{\nu} \biggr) \phi_{0} + \frac{R_{\nu}}{5} \Omega_{\rm B} - \frac{3}{5} \sigma_{\rm B}.
\end{equation}
The same strategy can be applied to more specific cases, such as the one where the scale factor interpolates between 
a radiation-dominated phase and a matter-dominated phase, as discussed in Eq. (\ref{interpolation}). In this case 
the solution of Eq. (\ref{monopole}) will be more complicated but always analytically tractable.
 
\renewcommand{\theequation}{6.\arabic{equation}}
\section{Concluding remarks}
\setcounter{equation}{0}

In this paper a systematic treatment of scalar perturbations has been discussed in the presence of a fully inhomogeneous 
magnetic field. No specific configuration has been assumed and the results are, in this sense, rather general.
Large-scale magnetic fields have been described in a fully consistent one-fluid MHD approach in curved 
space-time, which is particularly 
suitable for the analysis of  the low-frequency  part of the plasma spectrum. 
In this approach the charged components of the plasma (baryons and electrons) are 
in thermal equilibrium at a common temperature. The neutral components of the plasma (photons, neutrinos and 
CDM particles) are not affected directly by the presence of the magnetic fields. However, since 
magnetic fields gravitate and appear in the perturbed Einstein equations, also the 
initial conditions for the evolution of  the neutral species are modified in a specific fashion.
Since in MHD the Ohmic current is solenoidal, the baryon evolution equation is 
affected by a reduced Lorentz force term whose characteristic form is well known in flat-space MHD.
The combination of these different effects leads to a set of initial conditions 
for the CMB anisotropies, which is  rather different from the standard adiabatic (or isocurvature) modes. 

The main new  results of the present analysis can be summarized as follows:
\begin{itemize}
\item{} the problem of initial conditions for magnetized CMB anisotropies has been 
solved both in the conformally Newtonian gauge (more useful for theoretical calculations) 
 and in the synchronous gauge (more appropriate for numerical discussions);
\item{} if the curvature fluctuations are  adiabatic, magnetic fields 
modify the conventional adiabatic mode in a computable way and, as a consequence, the 
whole system of initial conditions  for CMB anisotropies is modified;
\item{} for the modified adiabatic mode, deep outside the horizon, the density contrasts
still satisfy the adiabaticity condition, however, as the horizon is crossed, a small 
non-adiabatic component develops;
\item{} if the fluctuations are isocurvature from the beginning, magnetic fields modify quantitatively
their amplitude, as shown in the case of the baryon isocurvature mode;
\item{}  the tight-coupling expansion has been revisited in the presence 
of a fully inhomogeneous magnetic field; 
\item{} it has been shown that,  because of the reduced Lorentz
force, both the zeroth and first order in the tight-coupling expansion are modified and this allows the initial conditions for the evolution of the brightness functions 
to be computed 
reliably   in both the adiabatic and isocurvature 
cases when fully inhomogeneous magnetic fields are present.
\end{itemize}
It is appropriate to conclude with some remarks concerning the interplay between large-scale 
magnetic fields and CMB physics. In various investigations, limits on the magnetic field
intensity are obtained on the basis of CMB considerations. Most of the times these limits 
refer to specific configurations. These discussions are certainly valuable; however, the
limits obtained should always be associated with a specific set of analytical initial 
conditions. As far as scalar fluctuations are concerned, this was not done up to now.
An example is Ref. \cite{koh} where no specific analysis of the initial condition problem was 
presented, but 
rather stringent limits were claimed. 

We must bear in mind that the analysis of CMB anisotropies requires the determination of a 
sizeable set of parameters. If the magnetic field were the only unknown, the task 
would be easier. However, this is unfortunately not the case. Hence, even within a specific 
magnetic field configuration, the limits should always refer to the specific set of initial conditions 
assumed in the analysis, as is normally done in the case when magnetic fields are absent.
We do hope that  future analyses along these directions may take the results 
of the present paper as a useful format for the systematic investigation of the effects of large-scale 
magnetic fields on the CMB anisotropies.

\newpage 
\begin{appendix}

\renewcommand{\theequation}{B.\arabic{equation}}
\setcounter{equation}{0}
\section{Boltzmann hierarchy}

As discussed in the bulk of the paper neutrinos are collisionless particles. In the following we are going to discuss 
the collisionless Boltzmann equation following the notation discussed in the present paper. 
The phase-space distribution of massless particles (massless neutrinos or photons) is defined to be 
\begin{equation}
f(x^{i}, P_{i},\eta) d x^{1} dx^{2} d x^3 d P_{1} d P_{2} d P_{3} = d {\cal N},
\label{f}
\end{equation}
giving the number of particles in a differential volume of phase space.
The function appearing in Eq. (\ref{f})  is scalar under canonical transformations, $P_{i}$ is the conjugate momentum. 
To derive and use the Boltzmann equation in curved space-time it is more convenient to work directly with the 
modulus $q$ and the direction $n_{i}$ of the 
 comoving three-momentum $q_{i}$ (where $ q_{i} = q n_{i}$ with $ n_{i} n^{i} = 1$).  
Denoting the longitudinal degrees of freedom of the perturbed geometry 
as in Eq. (\ref{longgauge}) and adopting the metric signature of Eq. (\ref{metric}), the 
relation between the comoving three-momentum and the components of the conjugate 
momentum are $ P_{i} = - q_{i} ( 1 -\psi)$ and $P_{0} = q ( 1 +\phi)$,

The total variation of the distribution function can be written as 
\begin{equation}
\frac{D f}{D \eta} = \frac{\partial f}{\partial \eta} + \frac{\partial x^{i}}{\partial\eta} \frac{\partial f}{\partial x^{i}} + 
\frac{\partial f}{\partial q} \frac{\partial q}{\partial \eta} + \frac{\partial f}{\partial n_{i}} \frac{\partial n^{i}}{\partial \eta} = 
\biggl( \frac{\partial f}{\partial \eta} \biggr)_{\rm coll},
\label{bz1}
\end{equation}
where the collisional term has been kept for later convenience, but it is zero in the case of neutrinos.
The collisionless part of Eq. (\ref{bz1}) can be perturbed around a configuration of thermal equilibrium by splitting the phase-space 
distribution as 
\begin{equation}
f(x^i, P_{j},\eta) = f_{0}(q) [ 1 + f^{(1)}( x^i, q, n_{j}, \eta)],
\label{split}
\end{equation}
where $ f_{0}(q)$ is the unperturbed phase-space distribution, which only depends upon the 
comoving three-momentum.

Inserting Eq. (\ref{split}) into Eq. (\ref{bz1}), only the terms that are first 
order will be kept. Since  the unperturbed phase-space distribution only 
depends upon the comoving three-momentum, Eq. (\ref{bz1}) becomes:
\begin{equation}
\frac{\partial f^{(1)}}{\partial \eta} +  n^{i} \frac{\partial f^{(1)}}{\partial x^{i}} + 
\frac{\partial \ln{f_0}}{\partial\ln q} [\psi' -  n_{i} \partial^{i} \phi] = 
\frac{1}{f_{0}}\biggl( \frac{\partial f}{\partial \eta} \biggr)_{\rm coll},
\label{bz2}
\end{equation}
where the geodesic equation has been used in order to obtain the expression of the 
time derivative of $q$ to first-order in the amplitude of the metric perturbations, i.e. 
\begin{equation}
\frac{d q }{ d\eta} = q \psi' - q n_{i} \partial^{i} \phi.
\end{equation}
Up to this point the derivation is valid  for both photons and massless neutrinos.

\subsection{Massless neutrinos and photons }

Going now to Fourier space and defining the 
reduced phase-space distribution for massless neutrinos as 
\begin{equation}
{\cal F}_{\nu}( \vec{k}, \hat{n}, \eta) = \frac{\int q^{3} d q f_{0} f^{(1)}}{\int q^{3} d q f_{0}},
\label{reddist}
\end{equation}
Eq. (\ref{bz2}) becomes, in the absence of collision term,
\begin{equation}
\frac{\partial {\cal F}_{\nu}}{\partial \eta} + i k\mu {\cal F}_{\nu} = 4 (\psi' - i k \mu \phi) ,
\label{bz3} 
\end{equation}
where $ \mu= \hat{n} \cdot\hat{k}$. The factor $4$ appearing in Eq. (\ref{bz3}) follows 
from the explicit expression of the Fermi-Dirac phase-space distribution and observing
\begin{equation}
\int q^{3} d q \frac{\partial \ln{f_{0}}}{ \partial \ln{q}} = - 4  \int q^{3} d q f_{0}.
\label{4fact}
\end{equation}
The reduced phase-space distribution can be expanded in series of Legendre 
polynomials as 
\begin{equation}
{\cal F}_{\nu}( \vec{k}, \hat{n}, \eta) = \sum_{\ell} (-i)^{\ell} ( 2 \ell + 1) F_{\nu\ell}(\vec{k},\eta) P_{\ell}(\mu).
\label{expF}
\end{equation}
Equation (\ref{expF}) will now be inserted  into Eq. (\ref{bz3}). 
The orthonormality relation for Legendre polynomials,
\begin{equation}
\int_{-1} ^{1}  P_{\ell}(\mu) P_{\ell'}( \mu) = \frac{2}{2 \ell + 1} \delta_{\ell\ell'},
\label{norm}
\end{equation}
together with the well-known recurrence relation 
\begin{equation}
(\ell + 1) P_{\ell + 1}(\mu) = (2 \ell +1) \mu P_{\ell}(\mu) - \ell P_{\ell -1}(\mu)
\label{rec1}
\end{equation}
allows to get a hierarchy of equations to be obtained for the various multipole moments. The procedure 
is to take the various moments of both sides of Eq. (\ref{bz3}). In doing so,
expressions like 
\begin{equation}
ik \int_{-1}^{1} \mu P_{\ell'}(\mu) {\cal F}_{\nu} d\mu = 2 i k \biggl[ (-i)^{\ell' + 1} \frac{\ell' +1}{ 2 \ell' + 1}  F_{\nu(\ell' +1)}  
+ (-i)^{\ell' -1}\frac{\ell'}{2\ell' + 1} F_{\nu (\ell' -1)}\biggr],  
\end{equation}
will appear; they  can be evaluated by using Eqs. (\ref{norm}) and (\ref{rec1}).
The full form of the Boltzmann hierarchy is then 
\begin{eqnarray}
&& {\cal F}_{\nu 0}' = - k {\cal F}_{\nu 1} + 4 \psi',
\label{mom0}\\
&& {\cal F}_{\nu 1}' = \frac{k}{5} [ {\cal F}_{\nu 0} - 2 {\cal F}_{\nu 2}] + \frac{4}{3} k \phi,
\nonumber\\
\label{mom1}
&& {\cal F}_{\nu\ell}' = \frac{k}{2\ell +1} [ \ell {\cal F}_{\nu,(\ell-1)}  - (\ell+1) {\cal F}_{\nu (\ell+1)}].
\label{mom2}
\end{eqnarray}
Equation (\ref{mom2}) holds for $\ell \geq 2$. 

The components of the energy-momentum tensor can be connected with the monopole 
 and dipole of the distribution function 
\begin{equation}
T_{\mu}^{\nu} = - \int \frac{d^{3} P}{\sqrt{-g}} \frac{P_{\mu}P^{\nu}}{P^{0}} f(x^{i}, P_{j}, \eta).
\end{equation}
Recalling 
the connection between conjugate momenta and comoving three-momenta we get, to zeroth order:
\begin{equation}
\rho_{\nu} = \frac{1}{a^4} \int d^{3} q q  f_{0}(q),
\end{equation}
while to first order the density contrast (monopole), the peculiar velocity field (dipole) and the 
quadrupole become
\begin{eqnarray}
&& \delta_{\nu} = \frac{1}{4\pi} \int d \Omega {\cal F}_{\nu}(\vec{k},\hat{n},\eta) = {\cal F}_{\nu 0},
\label{def1}\\
&& \theta_{\nu} = \frac{3i}{16\pi} \int d\Omega (\vec{k}\cdot \hat{n}) F_{\nu}(\vec{k}, \hat{n},\eta) = \frac{3}{4} k {\cal F}_{\nu 1},
\label{def0}\\
&& \sigma_{\nu} = -\frac{3}{16\pi }\int d\Omega \biggl[ (\vec{k}\cdot \hat{n})^2 - \frac{1}{3}\biggr] F_{\nu}(\vec{k},\hat{n},\eta) = 
\frac{{\cal F}_{\nu 2}}{2}.
\label{def2}
\end{eqnarray}
Inserting Eqs. (\ref{def1}) and (\ref{def2}) into Eqs. (\ref{mom0})--(\ref{mom2}), the system following from 
the perturbation of the covariant conservation equations can be partially recovered 
\begin{eqnarray}
&& \delta_{\nu}' = - \frac{4}{3} \theta_{\nu} + 4\psi',
\label{mom4a}\\
&& \theta_{\nu}' = \frac{k^2}{4} \delta_{\nu} - k^2 \sigma_{\nu} + k^2 \phi,
\label{mom4b}\\
&& \sigma_{\nu}' = \frac{4}{15} \theta_{\nu} - \frac{3}{10} k {\cal F}_{\nu 3},
\label{mom4}
\end{eqnarray}
with the important addition of the quadrupole (appearing in Eq. (\ref{mom4a})) and of the whole 
Eq. (\ref{mom4}), which couples the quadrupole, the peculiar velocity field, and the octupole ${\cal F}_{\nu3}$.
Equation (\ref{mom4}) is important. After neutrino decoupling, when initial conditions are set,
${\cal F}_{\nu 3}=0$.

Equation (\ref{bz2}) can also be made explicit in the case of photons. In order to  do so the collision term should be specified.
Before writing down the Boltzmann equation for the photons it is useful, in order to match with the standard notations, to define 
the photon brightness perturbation, conventionally denoted by $\Delta$:
\begin{equation}
f( x^{i}, q, n_{j}, \eta) = f_{0}\biggl( \frac{q}{1 + \Delta}\biggr),
\label{bright}
\end{equation}
where $f_{0}(q)$, i.e. the unperturbed phase-space distribution, denotes now the Bose-Einstein distribution.
Comparing Eq. (\ref{split}) with the expression given in Eq. (\ref{bright}) and expanding 
for $\Delta < 1$, the two definitions are connected as 
\begin{equation}
\Delta = - f^{(1)} \biggl(\frac{\partial \ln{f_0}}{\partial \ln{q}} \biggr)^{-1}, \,\,\,\,\,\,\, 
{\cal F}_{\gamma} = - \Delta \frac{\int q^{3} d q \frac{\partial \ln{f_{0}}}{\partial \ln{q}}}{\int q^{3} d q f_{0}} = 4 \Delta, 
\label{identities}
\end{equation}
where ${\cal F}_{\gamma}$ is the reduced phase-space distribution for the photons defined 
in full analogy with the case of the massless neutrinos, i.e. Eq. (\ref{reddist}). The second identity of Eq. (\ref{identities}) 
follows from the first and from Eq. (\ref{4fact}). Using the relation between the brightness 
perturbation and the phase-space distribution, the collisionless part of the Boltzmann equation can be 
written as 
\begin{equation}
\Delta' + i k\mu ( \Delta + \phi) = \psi'.
\end{equation}
The brightness perturbation, as defined in Eq. (\ref{bright}),  corresponds physically to a perturbation in the first Stokes parameter 
(conventionally denoted by ${\rm I}$), i.e. to the sum of the squared amplitudes of the radiation field. 
By specifying the collision term, the full form of the 
Boltzmann equation for photons can be written as 
\begin{eqnarray}
&&\Delta_{\rm I}' + i k\mu \Delta_{\rm I} = ( \psi' - i k \mu \phi) + \tau'\biggl[ - \Delta_{\rm I} + \Delta_{{\rm I}0} +  \mu v_{\rm b} - 
\frac{1}{2} P_{2}(\mu) S_{0}\biggr],
\label{DI}\\
&& \Delta_{\rm Q}' + i k\mu \Delta_{\rm Q} = \tau'\biggl[ - \Delta_{\rm Q} + \frac{1}{2}( 1 - P_{2}(\mu))S_0\biggr],
\label{DQ}\\
&& \Delta_{\rm U}' + i k\mu \Delta_{\rm U} = -\tau' \Delta_{\rm U}.
\label{DU}
\end{eqnarray}
Concerning Eqs. (\ref{DI})--(\ref{DU}) a few comments are in order. The brightness functions $ \Delta_{\rm Q}$ and 
$\Delta_{\rm U}$ correspond, respectively, to the Stokes parameters ${\rm Q}$ and ${\rm U}$.  The parameter 
${\rm Q}$ being the difference of the squares 
of the amplitudes of the radiation field, is sensitive to the linearly polarized radiation. Since ${\rm Q}$ and ${\rm U}$ (unlike ${\rm I}$ and ${\rm V}$)
change under rotations, once ${\rm Q}$ is included also ${\rm U}$ must follow. 
The source term appearing in Eq. (\ref{DI}), $S_0$ is, within our 
conventions:
\begin{equation}
S_{0} = \Delta_{{\rm I}2} + \Delta_{{\rm Q}0} + \Delta_{{\rm Q}2},
\end{equation}
where, with obvious notations, $ \Delta_{{\rm I}\ell}$ denotes the $\ell$-th multipole of $\Delta_{\rm I}$ and
similarly for $\Delta_{Q}$. The peculiar velocity field for the baryons has been written as 
\begin{equation}
v_{\rm b}  = \frac{\theta_{\rm b}}{i k}.
\label{defvel}
\end{equation}
The notation $v_{\rm b}$ is preferred here in view of the application to the 
tight-coupling expansion of the Boltzmann hierarchy for the photon brightness.
Finally, $\tau' = x_{e} n_{e} \sigma_{\rm T}\frac{a}{a_0}$ is, as previously introduced, the 
differential optical depth for Thompson scattering.
The evolution equation for $v_{\rm b}$ can be easily obtained, for instance by Fourier transforming Eq. (\ref{T1}) 
and by using Eq. (\ref{defvel}):
\begin{equation}
 v_{\rm b}' + {\cal H} v_{\rm b} + i k\phi  + \frac{\tau'}{\alpha}\biggl( 3 i \Delta_{{\rm I}1} + v_{b} \biggr) =  - \frac{i}{4\pi k }  
\frac{F_{\rm B}(k)}{ a^4 \rho_{\rm b}},
\label{vb}
\end{equation}
having defined 
\begin{equation}
\alpha = \frac{3}{4} \frac{\rho_{\rm b}}{\rho_{\rm r}}.
\end{equation}

Equations (\ref{DI})--(\ref{DU}) and (\ref{vb}) are discussed in Section 4.
\end{appendix}

\end{document}